%% file: main.tex
\documentclass[10pt,journal,compsoc]{IEEEtran}
\input{settings}

\begin{document}



\title{Securing Cloud FPGAs Against Power Side-Channel Attacks: A Case Study on Iterative AES}

\author{Nithyashankari Gummidipoondi~Jayasankaran,~\IEEEmembership{Student~Member,~IEEE,}
Hao~Guo,~\IEEEmembership{Student~Member,~IEEE,}
Satwik~Patnaik,~\IEEEmembership{Member,~IEEE,}
Jeyavijayan (JV)~Rajendran,~\IEEEmembership{Senior Member,~IEEE,}
and\\~Jiang~Hu,~\IEEEmembership{Fellow,~IEEE}
\IEEEcompsocitemizethanks{\IEEEcompsocthanksitem N.\ G.\ Jayasankaran is with Qualcomm India,  
KA, 560066.\protect\\
E-mail: gjn@qti.qualcomm.com
\IEEEcompsocthanksitem  H.\ Guo, S.\ Patnaik, and J.\ Rajendran are with the Department
of Electrical and Computer Engineering, Texas A\&M University, College Station, 
TX, 77843.\protect\\
E-mail: \{guohao2019, satwik.patnaik, jv.rajendran\}@tamu.edu
\IEEEcompsocthanksitem J.\ Hu is with the Department
of Electrical and Computer Engineering and the Department
of Computer Science and Engineering, Texas A\&M University, College Station, 
TX, 77843.\protect\\
E-mail: jianghu@tamu.edu}}


\input{abstract}

\maketitle

\IEEEpeerreviewmaketitle

\input{section_I_intro}

\input{section_II_background}
\input{section_III_previous_works}
\input{section_IV_experimental_setup}

\input{section_V_our_attack}
\input{section_VI_A_aes_primitive_placements}
\input{section_VI_B_defense_with_other_logic}

\input{section_VI_C_active_fences}
\input{section_VII_discussion_inferences}

\input{section_VIII_conclusion}

\bibliographystyle{IEEEtran}
\bibliography{main}

\input{bios}


\end{document}

%% file: settings.tex
\ifCLASSINFOpdf
\else
\fi

 \usepackage{cite}
\IEEEoverridecommandlockouts
\usepackage{algorithmic}
\usepackage{graphicx}
\usepackage{textcomp}
\usepackage{xcolor}
\def\BibTeX{{\rm B\kern-.05em{\sc i\kern-.025em b}\kern-.08em
    T\kern-.1667em\lower.7ex\hbox{E}\kern-.125emX}}
\usepackage{booktabs} 
\usepackage[compact]{titlesec}
\usepackage{enumitem}
\setlist{noitemsep,topsep=0pt,parsep=0pt,partopsep=0pt}
\usepackage{float}
\usepackage{graphicx}
\usepackage[hang,raggedright, nooneline]{subfigure}
\usepackage[font={small, bf}]{caption}
\usepackage{url}

\usepackage{setspace}
\usepackage{esvect}
\usepackage{tabu}
\usepackage{pifont}
\usepackage{indentfirst}
\usepackage{bbding}
\usepackage{multirow}
\usepackage{multicol}
\usepackage{amsmath,amssymb,amsfonts}
\usepackage{scalerel}[2016/12/29]
\usepackage{algorithmic}  
\usepackage{graphicx}
\usepackage{textcomp}
\usepackage{xcolor}
\usepackage{wasysym}
\usepackage{soul}
\usepackage{mwe}
\usepackage[export]{adjustbox}
\usepackage{pgfplots}
\pgfplotsset{compat=1.17}

\def\BibTeX{{\rm B\kern-.05em{\sc i\kern-.025em b}\kern-.08em
    T\kern-.1667em\lower.7ex\hbox{E}\kern-.125emX}}

 \newcommand\KORed { \textcolor{red}{\large \ding{55}} } 
\newcommand\OKBlue { \textcolor{blue}{\Checkmark} } 

\newcolumntype{C}[1]{>{\centering\arraybackslash}m{#1}}
\newcommand{\tr}[1]{\textcolor{black}{#1}}

\usepackage{times}
{}
\newcommand{\blue}[1]{\textcolor{black}{#1}}

\newcommand{\tp}[1]{\textcolor{black}{#1}}
\definecolor{matisse}{HTML}{1F77B4}
\definecolor{raw_sienna}{HTML}{CB6843}
\definecolor{green_pale}{HTML}{808000}

%% file: abstract.tex
\IEEEtitleabstractindextext{%
\begin{abstract}

The various benefits of multi-tenanting, such as higher device utilization and increased profit margin, intrigue the cloud field-programmable gate array (FPGA) servers to include multi-tenanting in their infrastructure. However, this property makes these servers vulnerable to power side-channel (PSC) attacks. Logic designs such as ring oscillator (RO) and time-to-digital converter (TDC) are used to measure the power consumed by security critical circuits, such as advanced encryption standard (AES).  Firstly, the existing works require higher minimum traces for disclosure (MTD). Hence, in this work, we improve the sensitivity of the TDC-based sensors by manually placing the FPGA primitives inferring these sensors. This enhancement helps to determine the 128-bit AES key using 3.8K traces. Secondly, the existing defenses use ROs to defend against PSC attacks. However, cloud servers such as Amazon Web Services (AWS) block design with combinatorial loops.  Hence, we propose a placement-based defense. We study the impact of (i) primitive-level placement on the AES design and (ii) additional logic that resides along with the AES on the correlation power analysis (CPA) attack results. Our results showcase that the AES along with filters and/or processors are sufficient to provide the same level or better security than the existing defenses.

\end{abstract}

\begin{IEEEkeywords}
PSC attack, CPA attack, cloud FPGA security, ring oscillators (ROs), time-to-digital converters (TDC), minimum traces for disclosure (MTD)
\end{IEEEkeywords}}


%% file: section_I_intro.tex
\section{Introduction}
\label{sec:intro}

\subsection{\tp{Security Vulnerabilities in Cloud FPGAs}}
\label{sec:cloud}

\tp{Hardware accelerators based on the field programmable gate arrays (FPGAs) support parallel processing and have better performance and power efficiency than CPUs and GPUs, respectively.}
Owing to these performance benefits, they are best suited for deploying compute-intensive tasks such as big data analytics, real-time video processing, and genomics research~\cite{amazon_f1}.
Hence, to meet user demands, cloud service providers such as Amazon Web Services (AWS), Microsoft Azure, IBM Cloud, and Texas Advanced Computing Center (TACC) have replaced CPUs with FPGAs in their cloud infrastructures~\cite{amazon_f1,micro_azure, ibm_cloud_fpga,tacc}. 
The use of FPGAs in  cloud has dramatically improved their computing capabilities~\cite{amazon_f1,cloudfpga_server,avnet_fpga}.  \tp{However, an FPGA instance can be underutilized when allocated to a single user, leading to FPGA resource wastage.  Therefore, the cloud servers can deploy multi-tenanting to avoid this wastage. Multi-tenanting allocates each FPGA instance to more than one user leading to increased FPGA utilization and profit margin~\cite{mt_uses}. However, in hindsight, it} introduces several security vulnerabilities in cloud---the attackers leverage these vulnerabilities to perform several attacks, namely, power side-channel (PSC) attack~\cite{Suh18_power, cld_vulnerable, pda20, inside_job, Ramesh18, remote_inter_tahoori}, fault attack~\cite{tahoori_voltage, Krautter_Gnad_Tahoori_2018, provelengios_PDA, Mahmoud_timing19, osc19}, and covert channel~attack~\cite{giechaskiel19, gie_iccd19, giechaskiel-a, gie_asiaccs18}.

Researchers have proposed several works demonstrating PSC attacks on cloud FPGAs.
\tp{The works~\cite{cld_vulnerable} and~\cite{inside_job}, demonstrates the CPA attack on a 128-bit AES design. Likewise, the work}
\cite{Suh18_power} demonstrates a \tp{simple power analysis (SPA)} attack on RSA for a threat model considering cloud FPGAs. 
Similarly, there have been works proposing defenses to thwart PSC attacks on AES~\cite{active_fences,cpamap20,cld_vulnerable}. 
However, it is important to note that in order to perform any valid/accurate security assessment, it is imperative that we consider realistic scenarios of the underlying application and also have access to attack vectors that compromise the security guarantees of the underlying application.
\tp{Hence, we focus on addressing the challenges in the existing attack and defense works. }
The following section discusses \tp{these~challenges.}

\subsection{Challenges in PSC Attack on Cloud FPGAs}

Here, we discuss the specific challenges that we encounter for PSC attacks on cloud FPGAs.
\tp{The two \blue{primary} challenges in the existing attack \blue{techniques} are (i)~repeatability of the attack, and (ii)~the impact of sensor design and placements. These challenges are discussed next in detail.}

\tp{\textbf{Repeatability.}} To demonstrate \blue{the efficacy of any} defense technique, we need an attack technique that is 100\% repeatable, i.e., launching the attack every time should retrieve the entire key. 
If the attack is not repeatable, then the increase in MTD may be due to improper implementation of the attack technique and not due to the proposed defense technique. The previous works either provide only 42\% attack repeatability rate~\cite{cld_vulnerable} or do not study this aspect~\cite{inside_job}.
    
\tp{\textbf{The impact of TDC-based sensor's placement and noise.} The sensor's placements} and the impact of noise on the efficacy of the CPA attack results must be studied, as these factors lead to an increase in the MTD. 
The attack feasibility reduces as the MTD increases. 
The increased MTD mandates the attacker to have access to an increased number of power traces. \tr{In prior works the defender analyzes the impact of the relative position of the TDC sensor with respect to the crypto core (AES). Additionally, the manual placements of TDC sensors that have been discussed in~\cite{cpamap20} and~\cite{tahoori_fpga_voltage} focus on fine and coarse calibration settings. Here, the total path length of the TDC sensor can be made to have small or large variations, by fine and coarse tuning, respectively. However, in our work, as explained in Section 1.2 and Section 5, we analyze the impact of manually placing the FPGA primitives such as look-up tables (LUTs) and carry chain (CARRY4) that infers the TDC sensor. Further details regarding TDCs are discussed in Section~\ref{sec:our_attack}.}

\subsection{Challenges in PSC Defenses on Cloud FPGAs}

The challenges in the existing defenses are discussed next.  

    \tp{\textbf{Realistic scenario.}}  Crypto algorithms such as AES and RSA are a part of an SoC or wireless communication network. Nevertheless, there are no previous works that study the impact of additional circuits that reside along with the crypto cores.
    
    \tp{\textbf{Manual placement of FPGA primitives.}} The design implemented on the FPGA is inferred using FPGA primitives, such as look-up tables (LUTs), flip-flops, and carry chains. The impact of manually placing these primitives that infer the circuit-under-protection is not studied in the existing works.
    
    \tp{\textbf{Challenges in active fence defense.}} The existing cloud FPGA defense~\cite{active_fences} uses ROs to reduce the SNR of the power traces collected by the TDC-based sensors. However, ROs fasten the FPGA aging by creating hotspots~\cite{temporal_sca19, tahoori_voltage} and also induce faults, making it vulnerable to fault injection attacks~\cite{pda20}.   Additionally, cloud servers such as AWS blocks FPGA bitstream having combinatorial loops. Hence, implementing a defense using RO is not practically possible.
    
    \tp{\textbf{Defense thwarting only RO-based sensors.}} The current defenses~\cite{mitigate_vol_attacks21} can only detect ROs implemented in the attacker's logic. However, a successful CPA attack on AES is achieved by using TDC-based sensors~\cite{active_fences, cpamap20, cld_vulnerable}. Hence, there is a need to secure the victim's logic from TDC-based sensors.  Therefore, this work focuses on addressing the above challenges in the cloud FPGA domain. 

\subsection{Contributions and Organization of the Paper}
The contributions corresponding to each of the challenges in the attack and defense domains are listed below.

\begin{itemize}
    \item  \tp{The impact of junction temperature on the MTD is studied. Maintaining this temperature within a limit aids in achieving a repeatable attack.} This repeatability means launching the attack every time retrieves the correct 128-bit AES key.  
    \item   \tp{The sensitivity of the TDC-based sensors is improved} by analyzing the impact of dissimilar net delays in the~sensor~design.
    \item  \tp{The resilience to the CPA attack is evaluated in a realistic scenario.} The defender's logic consists of AES, other crypto cores such as MD5 and SHA256, from the MIT-II common evaluation platform (CEP)~\cite{mitcep} SoC platform. 
    \item  \tp{The impact of the placement of FPGA primitives inferring the crypto core is analyzed. } 
    \item  \tp{The realistic defense scenario considered in this work} neither affects the reliability of the FPGA nor induces voltage faults. 
    \item  \tp{The defense scenario considered also} thwarts CPA attack irrespective of the type of sensor used by the attacker.  
\end{itemize}

The paper is organized as follows. In Section~\ref{sec:background}, \tp{we explain the background details related to PSC attack, sensors used, and the methodology of the CPA attack.} In Section~\ref{sec:previous_attack_defenses}, we explain the previous work related to the PSC attack and its defenses in the cloud FPGA domain. The experimental setup used in this work is explained in Section~\ref{sec:experimental_setup}. Section~\ref{sec:our_attack} discusses our enhanced PSC attack and the corresponding attack results. Section~\ref{sec:defense} shares the different methodologies proposed by this work to secure cloud FPGAs against PSC attacks and their experimental results.  Finally, in Section~\ref{sec:discussion_and_inferences}, the inferences and conclusion from our experiments are explained.

%% file: section_II_background.tex
\section{Background}
\label{sec:background}

This section discusses \tp{FPGA fundementals, PSC attack, different sensors inferred using FPGA~primitives, and CPA attack methodology}.

\subsection{Understanding FPGAs}
\label{subsec:understanding_fpgas}

\begin{figure}[!t]
\centering
\includegraphics[trim=0.2in 0.2in 0.2in 0.1in,clip, width=0.6\columnwidth]{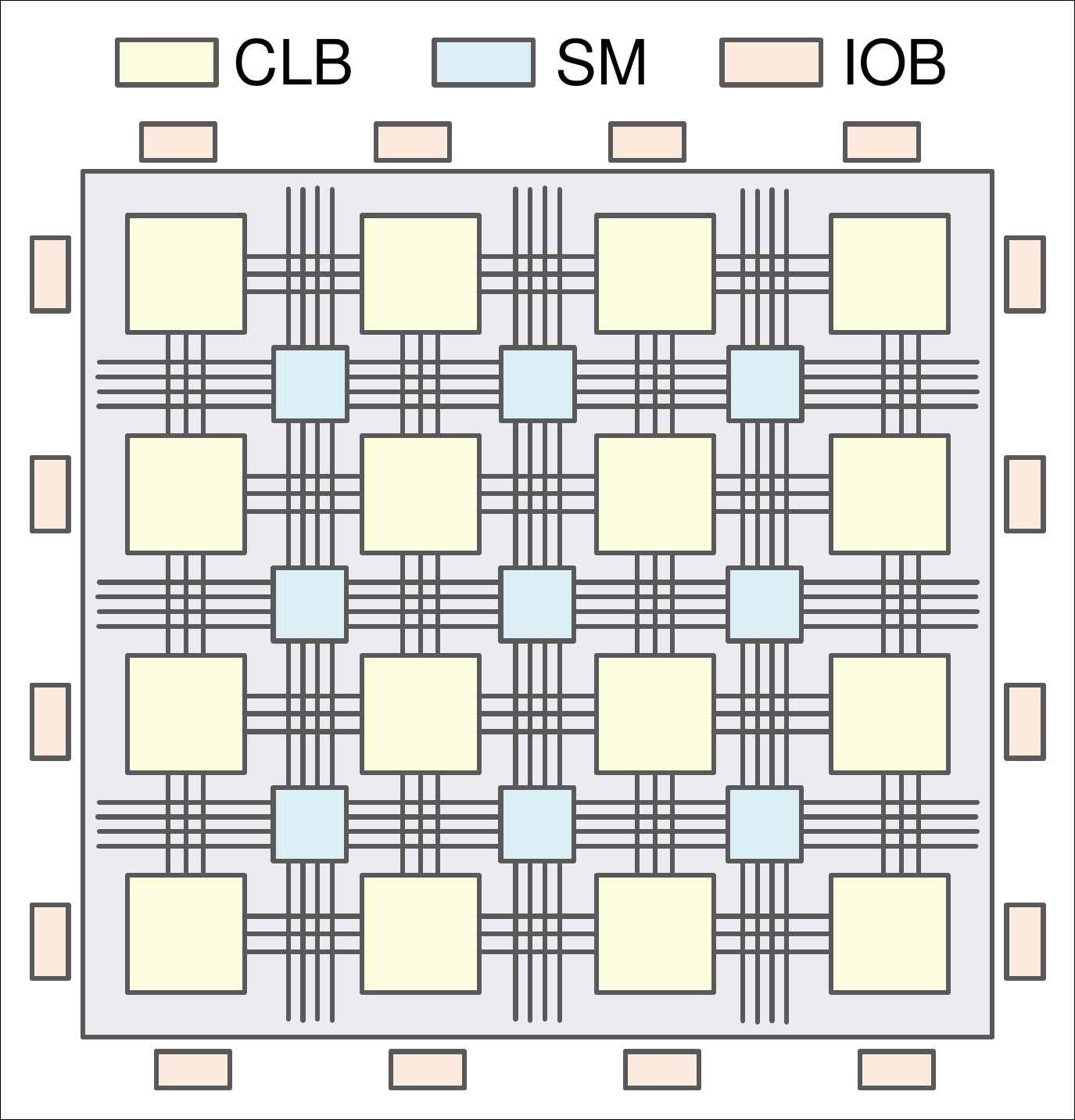}
\caption{\tp{Field-programmable gate array (FPGA) consists of configurable logic blocks (CLBs) connected to each other via switch matrix (SM). The FPGA communicates with the outside world using input-output blocks (IOBs).}}
\label{fig:fpga}
\end{figure}

\tp{ As this work uses Xilinx Zynq ZC706 FPGAs, we discuss the FPGA fundamentals using Xilinx terminologies. The FPGA consists of a two-dimensional array of configurable logic blocks. The communication between these logic blocks is via switch matrices, as illustrated in Fig.~\ref{fig:fpga}. These switch matrices consist of programmable interconnects that route signals. Based on the FPGA device used, each configurable logic block consists of a fixed number of slices. The FPGA used in this work has two slices per logic block. Each slice consists of the FPGA primitives such as LUTs, multiplexers, carry logics (CARRY4), and flip-flops. In this FPGA, each slice is composed of eight LUTs, three multiplexers, one CARRY4, and eight flip-flops. The LUTs, multiplexers, and CARRY4 implement combinational logic, whereas the flip-flops implement sequential logic. }

\subsection{PSC Attacks on Cloud FPGA}
\label{subsec:psa_cloud}

\tp{In a conventional PSC attack, the attacker has physical access to the FPGA boards, where cryptographic algorithms, such as AES, are implemented. For a given key and a plaintext, the AES generates a corresponding ciphertext. Using an oscilloscope connected to the power rail of the FPGA, the attacker measures the power consumed (power traces) during the execution of each step in the AES algorithm. First, the power traces are logged for different plaintexts and a fixed key. Following this, he/she runs the PSC attack on the collected power traces to determine the secret key~\cite{psa_dpa_attack}. Unlike the conventional setup, where the attacker has physical access to the FPGA board, he/she does not have this access in the cloud. Thus, in cloud FPGAs, the bitstream loading and debugging are done remotely. However, even without this access, researchers have successfully demonstrated PSC attack on the cloud FPGAs~\cite{Suh18_power, cld_vulnerable, pda20, inside_job, Ramesh18}. The following paragraph explains how the PSC attack is feasible even without physical access to the FPGAs on a cloud server.}

\tp{A successful PSC attack requires the sensing of voltage drop in the power distribution network (PDN) shared between the attacker and the victim logic. The PDN connects the supply voltage to the FPGA primitives, such as LUTs and flip-flops. Irrespective of the change in load current to these primitives, the supplied voltage must be constant. However, due to the PDN structure's asymmetry, the supply voltage drops in the presence of load current variations~\cite{Suh18_power, cpamap20, inside_job, cld_vulnerable}.  This variation in the load current is due to the data-dependent operations executed by the cryptographic algorithms.  That is, the voltage drop on the PDN reflects the power consumed by these algorithms.  The work~\cite{Suh18_power} utilizes this relationship to successfully perform the simple power analysis attack on the 1024-bit  Rivest, Shamir, Adleman (RSA) algorithm implemented on a cloud FPGA. Thus, even in the absence of physical access, the attacker uses a ring oscillator (RO) or a time-to-digital converter (TDC) circuits constructed using LUTs and flip-flops to measure the voltage drop rather than using an oscilloscope. The following section details the different sensors used in previous works launching PSC attack on cloud FPGA.}

\subsection{Sensors Used in Cloud FPGAs}
\label{subsec:sensors}
\tr{The most commonly used sensors for performing the PSC attack on cloud FPGAs are ROs~\cite{Suh18_power, Ramesh18} and TDCs~\cite{cpamap20, inside_job, cld_vulnerable}.} These sensors sense the voltage drop during the execution of the cryptographic algorithms. This sensing is feasible, as the voltage supplied to the logic gates has an inverse proportionality impact on the propagation delay of these gates~\cite{pant_des}. Therefore, a change in the propagation delay reflects the voltage drop. We now explain the different sensors using which the attackers measure the voltage drops in the FPGAs.

\begin{figure}[!b]
\centering
\includegraphics[trim=0.2in 0.28in 0.2in 0.3in,clip, width=0.65\columnwidth]{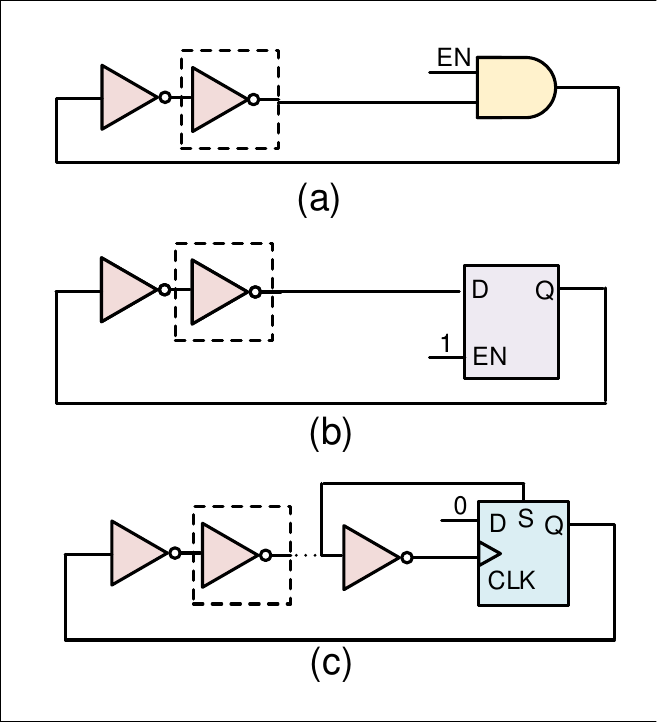}
\caption{Ring oscillator (RO) variants used in the prior works~\cite{Suh18_power, giechaskiel19_fpl, giechaskiel19, Ramesh18, Mahmoud_timing19, temporal_sca19, tahoori_fpga_voltage, tahoori_voltage, tahoori18_electrical, mitigate_vol_attacks21}. The dotted box corresponds to an even number of inverters. (a) A simple RO based on a combinatorial loop. A single LUT infers each of these inverters in the FPGA. (b) RO based on the latch, which avoids the formation of a combinatorial loop. (c) Another RO variant, based on flipflops that avoids the formation of combinatorial loops. This variant can escape Amazon AWS firewall filters that block FPGA bitstreams with combinatorial loops.}
\label{fig:old_sensor_designs}
\end{figure}

\subsubsection{Ring Oscillators (ROs)}

\tr{The ROs are widely implemented in cloud FPGAs to aid PSC attacks~\cite{Suh18_power, Ramesh18}, thermal covert-channel attacks~\cite{temporal_sca19}, voltage covert-channel attacks~\cite{tahoori_fpga_voltage, giechaskiel19, giechaskiel19_fpl, gie_asiaccs18}, and attacks based on inducing timing constraints violations.} The conventional ROs are inferred using the LUTs. Hence, when the power consumption around these LUTs changes, the voltage drop varies. This variation gets reflected as a change in the propagation delay in these LUTs~\cite{pant_des}. Therefore, the change in this propagation delay affects the time period and hence, the frequency of oscillations of the ROs. These ROs introduce several vulnerabilities listed below in the cloud FPGAs:
\begin{itemize}
\item In~\cite{temporal_sca19}, ROs are used as heat generators to aid the formation of thermal covert channels. As multiple users share a single FPGA, this work shows that the heat generated by one user can be observed by another using the same FPGA instance at a different time.
\item The ROs generate severe voltage fluctuations in~\cite{tahoori_voltage}. These fluctuations crash the FPGA within a few microseconds. As a result, the system can be used only after power-cycling.
\item They support voltage covert channel implementation~\cite{tahoori_fpga_voltage}.

\item the attacker can induce \tr{timing constraint violations} in the defender logic using ROs~\cite{Mahmoud_timing19}. These faults are critical as they are temporary and hence, impossible to detect.
\end{itemize} 

As ROs introduce various vulnerabilities in cloud FPGAs, the cloud service providers such as AWS blocks FPGA bitstreams containing combinatorial loops from getting deployed~\cite{aws_combo}. However, researchers designed ROs using latches and flip-flops to escape this filter in the Amazon firewall~\cite{mitigate_vol_attacks21}, as illustrated in Fig.~\ref{fig:old_sensor_designs}(b) and~\ref{fig:old_sensor_designs}(c). 

\begin{figure}[!t]
\centering
\includegraphics[trim=0.2in 0.2in 0.2in 0.1in,clip, width=\columnwidth]{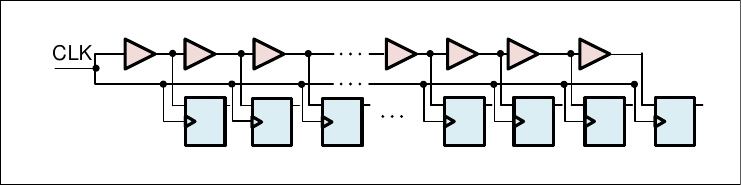}
\caption{Time-to-digital converter.}
\label{fig:tdc}
\end{figure}

\subsubsection{Time-to-Digital Converters (TDC)}
TDCs are also used in PSC attacks~\cite{cpamap20, inside_job, cld_vulnerable} to sense voltage drop variations. \tr{These sensors were originally developed to measure the propagation delay of logic gates (ASIC) and LUTs (FPGA)~\cite{Zick2013SensingNV, tdc_vernier_ro, Xia_tdc_6_6, wang_tdc_11}.} As the change in propagation delay reflects the change in voltage drop~\cite{pant_des}, the TDC design is leveraged to sense voltage variations. They can measure the time interval between two signal pulses or the arrival time of a single pulse and then convert it to a digital number. They are used in~\cite{tahoori18_electrical, mitigate_vol_attacks21} to measure voltage fluctuation in FPGAs. In~\cite{tahoori_fpga_voltage}, the TDC finds its application in the receiver side to sense the voltage change. Also, in~\cite{Mahmoud_timing19}, it is used as a voltage drop sensor. \tr{In this work, TDC-based sensors measure the dynamic power consumed by the 128-bit AES design.} 

We supply a clock signal to the chain of CARRY4 primitives, as shown in Fig.~\ref{fig:tdc}. Each CARRY4 primitive consists of four buffers connected in series, and the voltage drop experienced by these buffers influences the propagation delay experienced by them.  This delay, in turn, controls the number of buffers the clock signal travels through the chain. Finally, a latch controlled by the same clock signal gets connected to each buffer's output to log the number of buffers the clock has traveled. As TDCs can measure nanosecond delays and has higher accuracy~\cite{sensors_study}, they are used in measuring the power consumed by the AES design. Hence, in this work, we deploy TDCs to perform the same functionality. \tr{Apart from the impact of dynamic power consumption on the TDC sensor's output, this output is also affected by process variations and temperature. We have not studied the impact of process variation and temperature on the TDC output in the current work and will explore these impacts in our future works.} The following section explains the CPA attack methodology. 

\subsection{CPA Attack Methodology}
\label{subsec:cpa_attack}
\tr{This work uses the Hamming distance model based CPA attack, proposed in~\cite{cpa_attack}. A successful attack on a 128-bit AES design involves the following steps:}
\color{black}
\begin{enumerate}
\item Using the TDC-based sensors, the attacker measures the power consumed by the AES for encrypting different plaintexts. The number of power traces required by the attacker to determine the 128-bit key determines the strength of the attack --- the lesser the number of power traces stronger the attack is. 
\item For the first byte of the ciphertext, the attacker determines the output of the ninth round (first byte of intermediate ciphertext) using one out of 256 possible key guesses. He/She then calculates the Hamming distance between the first bytes of ciphertext and the intermediate ciphertext. Finally, the attacker repeats the Hamming distance calculation for the remaining 255 key guesses. 
\item The attacker then correlates each of the power traces measured in step 1  with the 256 different values of Hamming distances calculated from the previous step.  
\item The key guess corresponding to Hamming distance that has the maximum correlation with the power trace is the correct key~byte.
\item The first byte of the ninth round’s intermediate ciphertext and the key controls the first byte of ciphertext, i.e., the encryption of each byte of the intermediate ciphertext is independent of the other bytes. Hence, the attacker retrieves one key byte at a time. Therefore, the attacker repeats steps 2 to 4 for all the remaining 15 key bytes to determine the 128-bit AES key.
\end{enumerate}
\color{black}

%% file: section_III_previous_works.tex
\section{Previous Works}
\label{sec:previous_attack_defenses}

This section discusses the challenges in the previous works that proposed the PSC attack on 128-bit AES on a cloud FPGA platform and the defenses to thwart this attack.

\subsection{PSC Attack on AES Implemented on Cloud FPGAs}
Several works demonstrated the PSC attack on the AES implemented on the victim's side~\cite{inside_job, cld_vulnerable}. In~\cite{cld_vulnerable}, a TDC-based sensor is used to measure the power consumed by the AES for each encryption. This work can retrieve the 128-bit AES key \tp{using $500k$} power traces. However, as mentioned in this work, the attack success rate is only 42\%, i.e., the attack is not repeatable. Apart from demonstrating the PSC attack, another similar work~\cite{inside_job} shows the impact of the distance between the attacker and victim on the number of traces required to retrieving the AES key. This work requires $1.8k$ traces to retrieve one key byte.

\subsection{Defenses to Thwart PSC Attacks}
    
    \textbf{Active fences~\cite{active_fences}.} \tr{ This work proposes to reduce the signal-to-noise ratio (SNR) by increasing the noise in the AES encryptions. This reduction in SNR is achieved by enabling and disabling a group of ring oscillators (i) randomly and (ii) using the output value of the TDC-based sensor that senses the power consumption of the AES crypto under protection. With this defense, the MTD to determine one key byte is as high as $120k$ traces using the former method and $300k$ traces for the latter. In this technique, the ROs surrounds the crypto algorithms to form a fence. This implementation secures these algorithms against PSA.} As explained in the work, the RO  is implemented using a single look-up table (LUT) that infers a two-input NAND gate. The output of the NAND gate is fed back to one of the inputs forming the RO. The other input connects to the enable signal, which enables or disables the RO.
    
    \textbf{CPAmap~\cite{cpamap20}.} This work aims to understand the underlying mechanisms and dependencies of chip-internal side-channel attacks. It investigates the sensitivity of the TDC-based sensors on different locations on the FPGA exhaustively. It is achieved by running the correlational power analysis (CPA) attack on the traces collected by the sensors placed at different locations.  The impact of Vivado implementation settings on the bitstream generations is also studied. However, there are few challenges in this work, as mentioned below. 
    \begin{itemize}
        \item After executing the PSC attack at multiple locations on the FPGA, the cloud service provider identifies the locations on the FPGA that are not safe for the crypto algorithms. These are the locations where the sensor can determine the key with less MTD. The provider refrains from allocating these locations to any user leading to FPGA resource wastage. 
         \item The results of this work prove that unless an exhaustive experiment is conducted on each FPGA, the cloud service provider cannot determine the sensitive locations on the FPGA. This process is time-consuming, and it might have to be repeated as the chip ages, as aging affects the sensitivity of the FPGAs.
    \end{itemize}
    
    \textbf{Mitigating voltage attacks~\cite{mitigate_vol_holcomb}.} \tr{The work proposes a defense against fault attacks. It senses or identifies the source of voltage attacks (logics that cause abnormal voltage drops) on the cloud FPGAs  during runtime, i.e., when the attack is executed.} The attacker implements malicious ROs triggers sudden voltage drops causing fault attacks~\cite{tahoori_voltage}. To defend against this attack, the cloud service provider in~\cite{mitigate_vol_holcomb} deploys sensors at multiple locations on the FPGA. If the voltage drop measured is less than the pre-defined value, the clock to that region is disabled, thereby thwarting the attacks. \tr{Though this defense is to secure against fault attacks, as it deducts the presence of ROs it can be leveraged to secure against PSC attacks due to ROs~\cite{Suh18_power}}.

\tr{\textbf{Mitigating electrical-level attacks~\cite{mitigate_vol_attacks21}.} This work is an offline countermeasure that translates the FPGA bitstreams to flat technology mapped netlist. This netlist is then parsed for the presence of circuits such as ROs and TDCs. If these circuits are detected in the netlist, the bitstream is blocked from being deployed on the cloud servers, as they aid PSC attacks.} 

%% file: section_IV_experimental_setup.tex
\section{Evaluation Plan and Setup}
\label{sec:experimental_setup}

\begin{figure}[!t]
\centering
\includegraphics[trim=0.1in 0.25in 0.1in .25in, clip, width=0.35\textwidth]{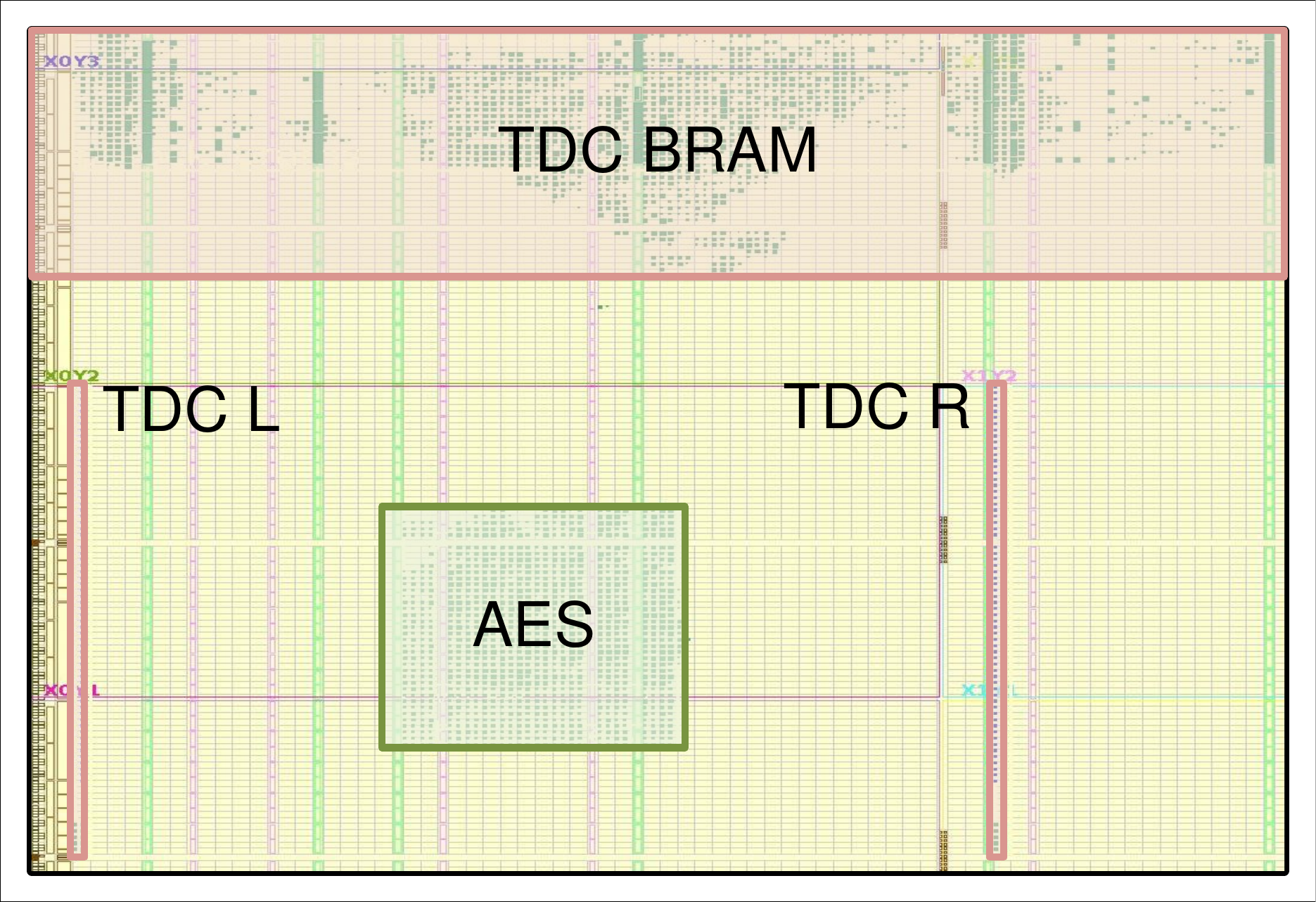}
\caption{FPGA device view of the power side-channel attack (PSA) setup. The victim's logic (green) is the 128-bit iterative AES crypto algorithm. The attacker uses two TDC-based sensors to measure the power consumption and the block RAMs (BRAMs) to store these power traces. The attacker's logic is shown in red. }
\label{fig:original_setup}
\end{figure}
This section discusses the evaluation plan of this work. This work focuses on (i) improving the PSC attack and (ii) studying the impacts of primitive-level placement and extra logic placed along with the crypto design.  Rather than proposing a new attack or defense technique, this work studies the impact of primitive-level placement in the sensor design (attack) and the crypto design under protection (defense). Also, we evaluate the CPA attack on multiple case studies where other logic designs such as processors and filters are placed along with the AES under protection.  Our baseline experimental setup consists of two TDC-based sensors, one on the left and one on the right of the AES, as illustrated in Fig.~\ref{fig:original_setup}.

\begin{figure}[!t]
\centering
\includegraphics[trim=0.1in 0.25in 0.1in .25in, clip, width=0.47\textwidth]{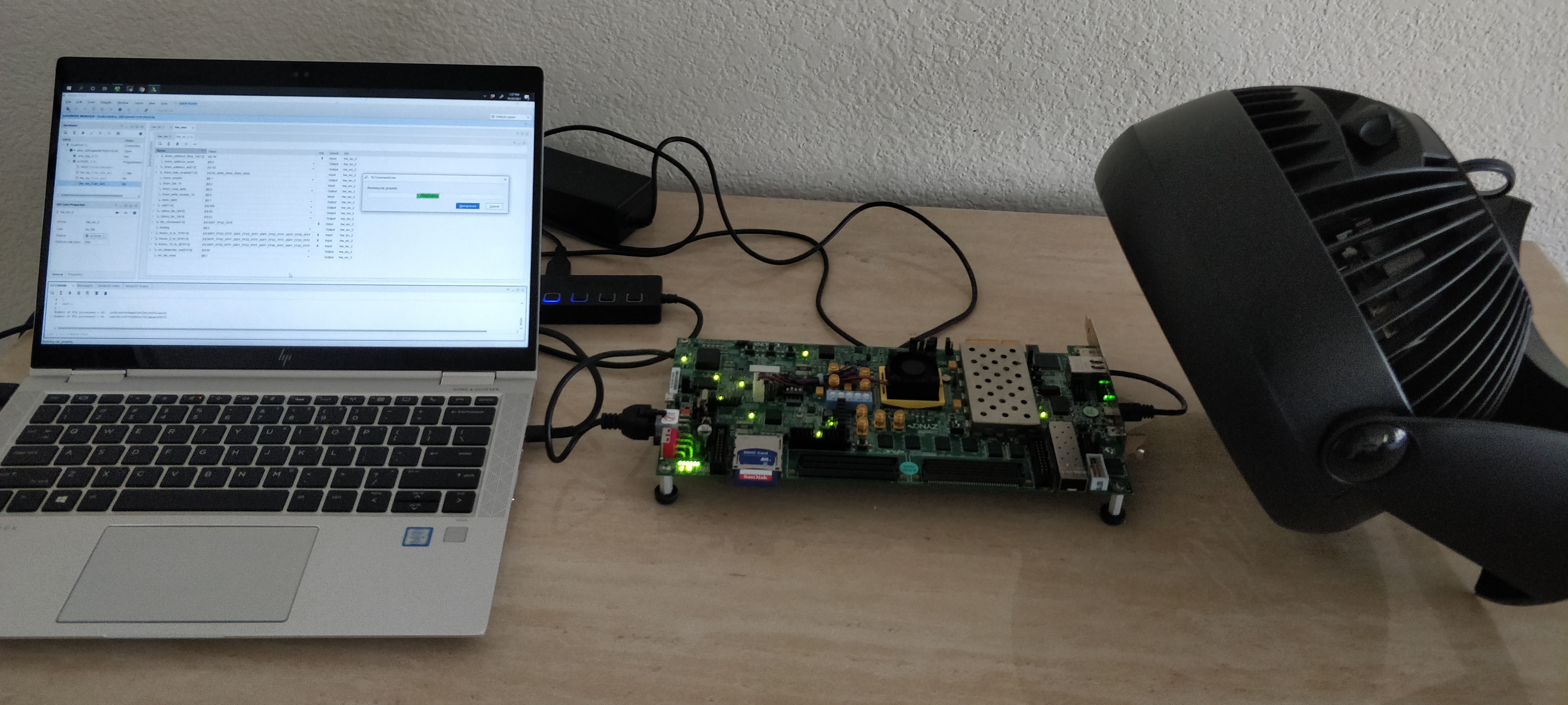}
\caption{\tp{All experiments in this work are demonstrated on the Zynq ZC706 FPGA evaluation platform. The board is connected to the host machine, where the power traces are collected to perform correlation power analysis (CPA) attacks. Apart from the fan on the heat sink to reduce the FPGA junction temperature, we also placed a table fan to cool the FPGA board further.}}
\label{fig:exp_setup}
\end{figure}

\subsection{Threat Model}
\label{subsec:threat_model}

This work assumes that the cloud services support multi-tenanting, i.e., multiple users can share one FPGA instance, in which one of the users can be an attacker (a malicious user). As the cloud services correspond to a common resource pool shared by multiple users, there is a high probability that the allocation of the victim's logic and the attacker's (malicious user) logic happens in the same FPGA instance.  Additionally, the attacker or defender does not have physical access to the FPGAs. Therefore, he/she cannot probe the power rails to measure the power consumed using oscilloscopes.

This work chooses the 128-bit iterative version of the AES crypto algorithm as the victim's logic, whose source code is available at~\cite{aes_code}. It also includes a virtual input output (VIO) debug IP core~\cite{vio_xilinx} to transmit and receive data from and to the AES core via the JTAG signals of the FPGA. The attacker's logic uses the TDC-based sensors to measure the power consumed by the AES encryption. Additionally, the attacker's logic includes block RAMs and a VIO debug IP core to store and send the power traces, respectively, to the user. The source code of the CPA attack~\cite{cpa_fei} is tailored to suit our attack setup. This work uses the Zynq ZC706 evaluation platform to execute all our experiments. \tr{This platform is unmodified and we have not removed or added any capacitors on the evaluation platform. As our threat model focuses only on remote PSC attacks, where the attacker does not have access to the physical pins of the FPGA or the capacitors associated with the power supply, the evaluation platform is used without any modifications for conducting all our experiments.} This evaluation platform along with the host machine and the cooling fan is shown in Fig.~\ref{fig:exp_setup}. The following section shares our attack contributions and the CPA attack results on our enhanced attack setup. Similarly, the CPA attack results, power, and area consumption of the different placement-based experiments and experiments with extra logic follow this section.

%% file: section_V_our_attack.tex
\section{Our Enhanced Attack on 128-bit AES}
\label{sec:our_attack}

This section focuses on setting up a repeatable attack to evaluate the resilience offered by our defense securing cloud FPGAs against PSC attack, i.e., the crypto key must be retrieved every time launching the attack on the cloud FPGAs. If the attack is not repeatable, then the increased minimum traces to disclosure (MTD) may be due to improper implementation of the attack itself and not the proposed defense. However, the previous works~\cite{cld_vulnerable, active_fences, cpamap20} either do not have a 100\% attack success rate or do not evaluate the success rate.  Hence, in this work, we propose fine-tuning the sensor design to reduce the MTD required to determine the 128-bit AES key and make the attack repeatable. The following sections discuss (i) the manual placement of FPGA primitives inferring the sensors, (ii) determining the time instant to which the CPA attack has the highest correlation, and (iii) studying the impact of junction temperature.

\begin{figure}[!t]
\centering
\includegraphics[trim=0.1in 0.3in 0.1in .25in, clip, width=0.35\textwidth]{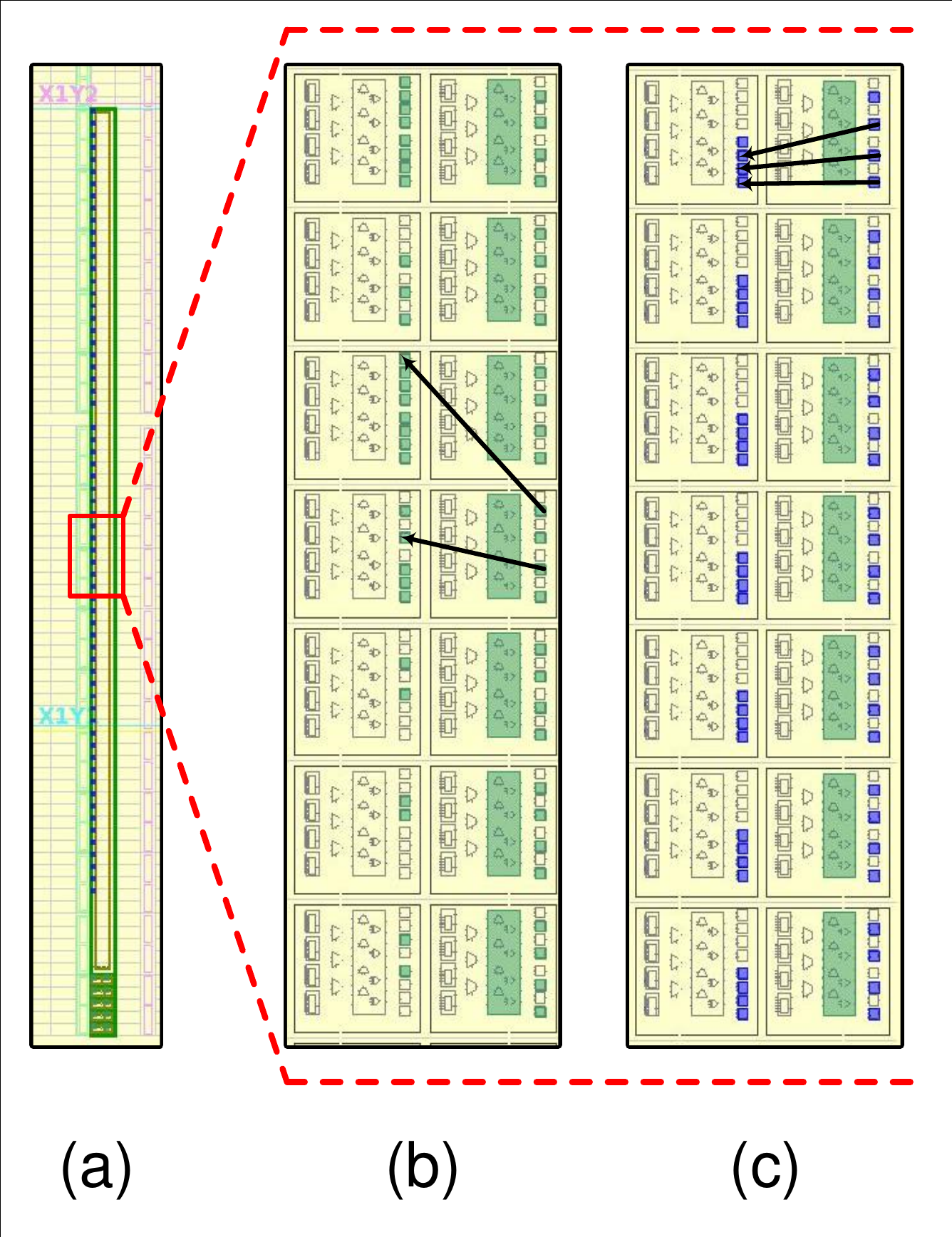}
\caption{(a) FPGA device view of the TDC-based sensor. The zoomed version of the section of sensor enclosed in the red box are shown in Fig.~\ref{fig:tdc_placements} (b) and (c). (b) The Vivado tool automatically places the FPGA primitives corresponding to the sensor, leading to unequal net delay between the different latch-flipflop pairs. (c) The attacker judiciously places the FPGA primitives such that there is an approximately equal net delay between the different latch--flip-flop pairs.}
\label{fig:tdc_placements}
\end{figure}

\begin{figure}[!t]
\centering
\includegraphics[trim=0.25cm 0 0.25cm 0, clip, width=0.37\textwidth]{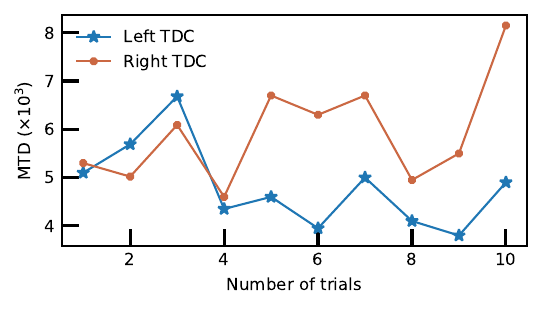}
\caption{{Correlation power analysis (CPA) attack results on the enhanced attack setup with two sensors, left and right TDCs. The reliability of the attack is tested by collecting the traces multiple times for the same build. In each trial, all 16 key bytes are retrieved successfully. Minimum number of traces for disclosure (MTD) and time to digital converter (TDC).}}
\label{fig:cpa_results_repeat}
\end{figure}

 \textbf{Sensor Placement.} \tr{In this work, we explore the impact of manually placing each FPGA primitive inferring the TDC-based sensor.  As explained in Section~\ref{sec:background}, the TDC consists of a chain of CARRY4 primitives. Each primitive consists of four outputs, where each output gets connected to a latch. These latches are, in turn, connected to the flip-flops. When Vivado places these primitives automatically, the net delay between the latches and their corresponding flip-flops are different for different pairs, as shown in Fig.~\ref{fig:tdc_placements} (b).  This difference impacts the sensor output. By manually placing the flip-flops (colored in blue) illustrated in Fig.~\ref{fig:tdc_placements} (c), the net delay is approximately equal between all the latch--flip-flop pairs. This manual placement of these latches and flip-flops is achieved by providing user-defined constraints, such as LOC and BEL, during bitstream generation~\cite{xdc_xilinx}.  With the help of this primitive-level placement, we could retrieve the 128-bit AES key using $10.7k$ and $10k$ traces using the left and right TDC-based sensors, respectively.} 

\textbf{Determining the Time Instant at which the Highest Correlation Occurs.} \tr{As explained in Section~\ref{sec:background},  in the CPA attack, the key guess corresponding to Hamming distance that has a maximum correlation with the power trace is the correct key byte. Here, the Hamming distance is calculated between the tenth and ninth round ciphertexts. Hence, rather than considering the power consumed during all the ten rounds of AES encryption, we consider only the power consumed by the flip-flops during the tenth round of encryption. As a result, it helps achieve a low MTD ($2.3k$ and $2.1k$ traces using the left and right TDC-based sensors, respectively) compared to the setup, where the power traces corresponding to all the ten rounds are considered ($10.7k$ and $10k$ traces using the left and right TDC-based sensors, respectively).} Additionally, it also reduces the number of samples collected for each encryption.
\tr{We have added this section to understand what MTD the attacker can achieve in a best-case scenario where he/she uses only the power traces corresponding to the ninth and tenth rounds. Our results show that the MTD is lesser when only these rounds are considered rather than all 10 rounds of power traces, i.e, the power traces corresponding to the other rounds act as noise and hence increase the MTD to determine the 128-bit key.}

In this work, by fine-tuning the sensor design, we propose to demonstrate a repeatable attack on the 128-bit AES algorithm implemented on the Zynq ZC706 evaluation board. As shown in Fig.~\ref{fig:cpa_results_repeat}, the 128-bit key is retrieved using $3.8k$ traces. Furthermore, on all the ten trials, the attack successfully determines the 128-bit key. As explained in Section~\ref{sec:our_attack}, the attack setup consists of the iterative AES algorithm and two TDC-based sensors, one on each side of AES. This setup is illustrated in Fig.~\ref{fig:original_setup}. The MTD reported by the previous works to determine the AES key is either around $1.8k$ traces~\cite{active_fences} for one key byte or $500k$ for 16 key bytes. However, as illustrated in Fig.~\ref{fig:cpa_results_repeat}, the minimum and maximum MTD to determine the 128-bit key are $3.9k$ and $8.1k$ traces, respectively. The manual placement of each primitive in the sensor design aids in reaching this low MTD. This placement ensures approximately equal net delays between each latch--flip-flop pair, thereby reducing the errors in the sensor output, reflecting the power consumed by the defender's logic.

 \textbf{Understanding the Impact of Junction Temperature.} Powering on the evaluation board for a long time increases the junction temperature of the FPGA. Additionally, the rate at which the different parts of the FPGA cool depends on these heat sink paths~\cite{xilinx_power_methodology}. Hence, the increased junction temperature induces temperature-dependent noise. This noise impacts the CPA attack resulting in increased MTD to determine the 128-bit key. As shown in Table~\ref{tab:cpa_results_org}, with increased junction temperature, the CPA attack requires a higher MTD to retrieve all key bytes. However, maintaining this temperature below 30$^{\circ}$C  (using a cooling fan), the CPA attack retrieves all key bytes within $6k$ traces. \tr{The increase in junction temperature is already mitigated by placing the evaluation board in the refrigerator in~\cite{cpamap20}. However, our objective to study the impact of junction temperture on the MTD is to show that we could achieve a repeatable attack.}

\begin{table}[!t]
    \centering
    \caption{{Correlation power analysis (CPA) attack results on our enhanced design. The design consists of AES (victim's logic) and the time-to-digital converter (TDC) (attacker's logic). The TDC is used for sensing the voltage variations in the victim's logic. The impact on temperature is studied. Minimum number of traces for disclosure (MTD).}}
    \resizebox{\columnwidth}{!}{
    \label{tab:cpa_results_org}
       {\tabulinesep=1mm
    \begin{tabu}{|c|c|c|c|c|c}
    \hline
{\bf Build }                      & \multicolumn{2}{|c|}{\textbf{No cooling time}} &   \multicolumn{2}{|c|}{\textbf{30 minutes cooling time}} \\\cline{2-5}
\textbf{number}                   & \textbf{bytes recovered} & \textbf{MTD}        & \textbf{bytes recovered} & \textbf{MTD} \\\hline\hline
                1                 & 16 & 5400  & 16 & 3900         \\\hline
                2                 & 15 & 8471  & 16 & 4600         \\\hline
                3                 & 15 & 5977  & 16 & 5100         \\\hline

    \end{tabu}}} 
\end{table}

\tr{Table~\ref{tab:comparison_attack} lists the existing works on PSC attack along with this work. The attack implemented in this work has 100\% repeatability. Additionally, by manually placing the sensor primitives we could achieve a very low MTD of 3800 traces compared to the sensor design that was automatically placed by the Vivado software that requires an MTD of $27k$ traces.  }

%% file: section_VI_A_aes_primitive_placements.tex
\section{Our Defense Analysis}
\label{sec:defense}
\begin{figure*}[!t]
\centering
\includegraphics[trim=0.2cm 0.2cm 0.2cm 0.2cm, clip, width=\textwidth]{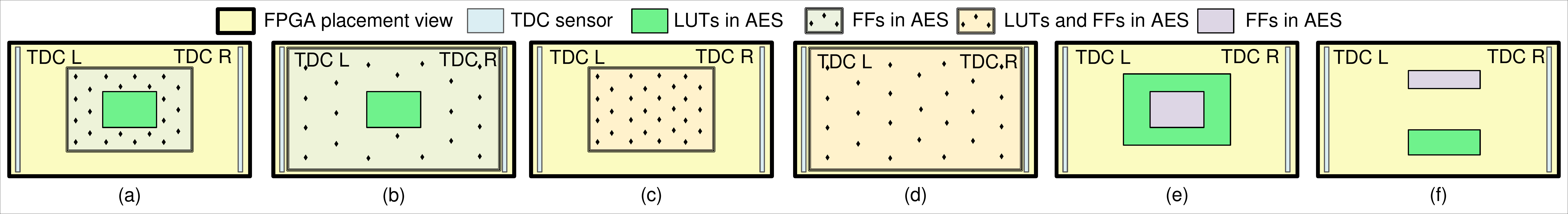}\hfill
\caption{Placement strategies. (a) The flip-flops in the AES design have three empty slices between each other along the X and Y axes. (b) The flip-flops in the AES design have six empty slices between each other along the X and Y axes. (c) The flip-flops and LUTs in the AES design have three and two empty slices, respectively between each other along the X and Y axes. (d) The flip-flops and LUTs in the AES design have six and four empty slices, respectively between each other along the X and Y axes.  (e) The LUTs surround the flip-flops in the AES design. (f) The flip-flop block (violet) and the LUT block (green) are placed one above the other.}
\label{fig:primitve_1}
\end{figure*}

\subsection{The Impact of Primitive-Level Placement of AES on CPA Attack}
\label{subsec:impact_of_primitive_level_placement_of_aes_on_cpa_attack}

 As given in Section~\ref{subsec:cpa_attack}, for the correct key guess, the correlation between the power trace and the Hamming distance is maximum. This Hamming distance corresponds to the number of bit-flips between the AES's ninth and tenth (ciphertext) round outputs. These bit-flips are the change in flip-flop outputs that are responsible for the power consumption.  Unlike the LUTs, the flip-flops (edge-sensitive) output changes at the rising edge of the clock. Thus, the flip-flops in the AES predominantly contribute to dynamic power consumption compared to the LUTs. As flip-flops mainly contribute to dynamic power consumption, we study the impact of the placement of these FPGA primitives manually on the CPA attack, rather than allowing the Vivado to perform automated placement of these primitives.  Vivado places the FPGA primitives corresponding to the same logic as close as possible to reduce the wirelength delay. However, in this work, to ensure that the sensors cannot sense all the flip-flops of the AES, the defender spreads out the flip-flops across the clock region. The following are the different placement strategies the primitives in the AES algorithm have been placed on the Zynq 7000 series FPGA. 

\begin{table}[!t]
    \centering
    \caption{\tr{The PSC attack results on different FPGA platforms reported by the works~\cite{cld_vulnerable, inside_job} and this work. Minimum number of traces for disclosure (MTD). Power side-channel (PSC) attack. Not applicable (NA). Evaluation platform (EP).}}
    \resizebox{0.9\columnwidth}{!}{
    \label{tab:comparison_attack}
       {\tabulinesep=0.8mm
    \begin{tabu}{|c||c|c|c|}
    \hline
\multirow{2}{*}{\bf Attack property} & {\bf PSC}                            & \textbf{Inside }                     & \multirow{2}{*}{\bf This work}\\
                                     & \textbf{attack\cite{cld_vulnerable}} & \textbf{job~\cite{inside_job}} &\\\hline\hline
\textbf{Repeatability}   & 42\%      & NA    & 100\%\\\hline
\textbf{MTD}             & $500k$    & 1800  & 3800\\\hline
\textbf{\# of key bytes} & \multirow{2}{*}{16}        & \multirow{2}{*}{Not available}      & \multirow{2}{*}{16}  \\
\textbf{retrieved} &         &      &  \\\hline
\textbf{Platform} & Ultrascale+     & XC6SLX75 & ZC706 EP \\\hline
    \end{tabu}} }  
\end{table}

\begin{enumerate}
\item{\textbf{Only the flipflops are spread out.}}
In this technique, only the flipflops in the AES design are spread across the clock regions, as illustrated in Fig.~\ref{fig:primitve_1}(a) and~\ref{fig:primitve_1}(b). The flops correspond to the plaintext and the key registers in the crypto algorithm. Increasing the number of empty slices between each flop along the X- and Y-axis of the FPGA achieves different spread levels. 

\item{\textbf{Both the flip-flops and LUTs are spread out.}}
In this technique, both the LUTs and flip-flops in the AES design are spread across the clock regions, as illustrated in Fig.~\ref{fig:primitve_1}(c) and~\ref{fig:primitve_1}(d). Increasing the number of empty slices between each flop and LUT along the X- and Y-axis of the FPGA achieves different spread levels. 

\item{\textbf{Flip-flop block surrounded by the LUT block.}}
Here, the flip-flops constrained within a block are surrounded by the LUTs, as shown in Fig.~\ref{fig:primitve_1}(e). 

\item{\textbf{Flip-flop block and the LUT block are placed one above the other.}}
Like the previous placement, the flip-flops and the LUTs are constraints to an individual block of regions on the FPGA and placed one above the other. Here, different spread levels are generated with an increasing number of empty slices between the two blocks, as illustrated in Fig.~\ref{fig:primitve_1}(f). 
\end{enumerate}

We now discuss the CPA attack results for these different placement strategies explained above.

\begin{figure*}[!t]
\centering
\includegraphics[trim=0 0.24in 0 0, clip, width=\textwidth]{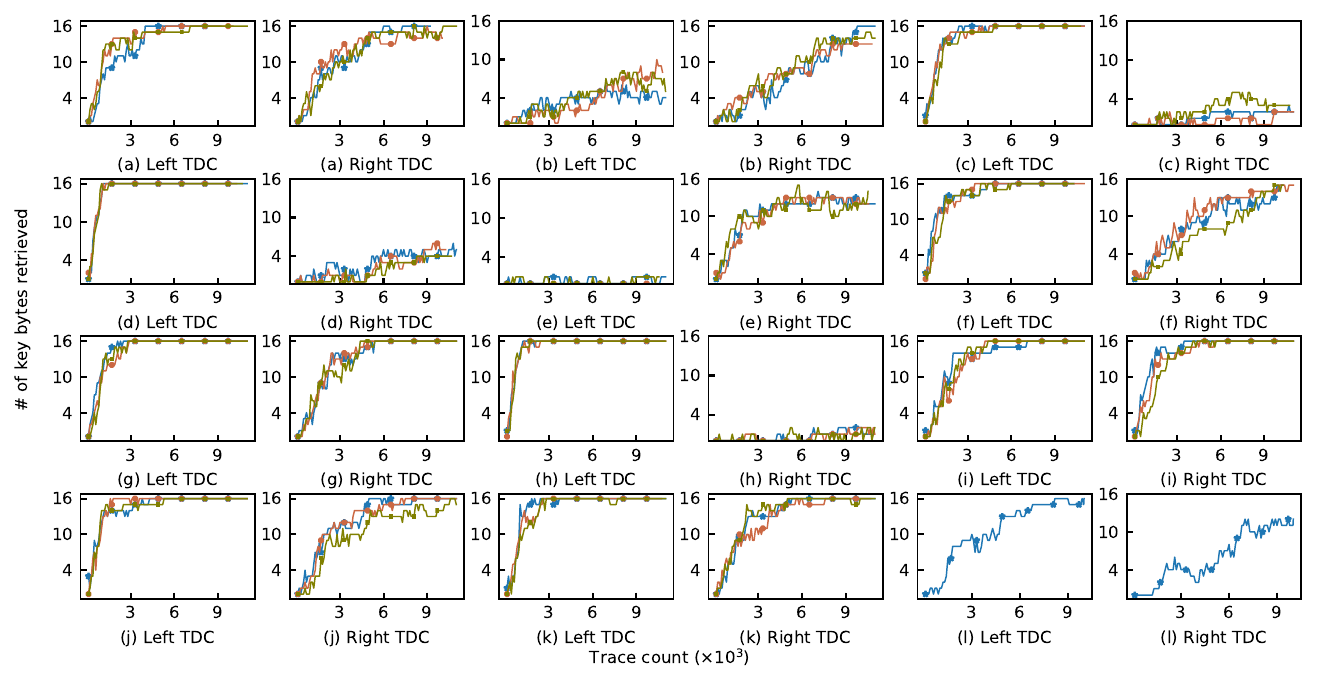}\hfill
\includegraphics[trim=0.16in 0.16in 0.16in 0.16in, clip, width=\textwidth]{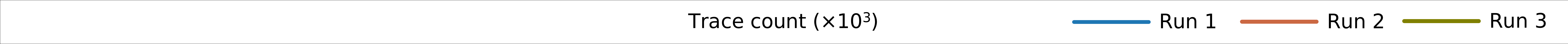}
\caption{ Correlation power analysis (CPA) attack on the power traces collected by the left and right TDC-based sensors. The flipflops in the AES design have (a) two, (b) three, and (c) six empty slices between each other along the X and Y axes. The flipflops and LUTs in the AES design have (d) three and two empty slices, (e) six and four empty slices, respectively, between each other along the X and Y axes. (f) The LUTs surround the flipflops in the AES design. The placement of the flipflop block is above the LUT block. There are (g) 22, (h) 75, and (i) 113 empty slices in between the blocks. The width and depth of the flipflop block are 30 and 4, respectively. The width and depth of the LUT block are 30 and 13, respectively.  There are (j) 75 and (k) 113 empty slices in between the blocks. The width and depth of the flipflop block are 90 and 2, respectively. The width and depth of the LUT block are 90 and 6, respectively.}
\label{fig:cpa_results_placement_spread}
\end{figure*}

\subsection{CPA Attack on Our Defense Case Studies}
\label{subsec:our_defense_results}

\subsubsection{Impact of Primitive-Level Placement of AES on the MTD}
\label{subsubsec:primitive_placement_results}

\textbf{Spreading the flipflops and LUTs.} The CPA results on the build based on the different primitive placement strategies discussed in Section~\ref{subsec:impact_of_primitive_level_placement_of_aes_on_cpa_attack} are illustrated in Fig.~\ref{fig:cpa_results_placement_spread}.
When there are two empty slices between any two flip-flops, most of the key bytes are retrieved using less than $5k$ traces, as illustrated in Fig.~\ref{fig:cpa_results_placement_spread}(a). Increasing the empty slices to three requires around $10k$ traces to determine 16 and 10 key bytes by the right and left sensors, respectively, as shown in Fig.~\ref{fig:cpa_results_placement_spread}(b). However, increasing the number of empty slices between the AES flip-flops has different effects on the left and right sensors.  

Fig.~\ref{fig:cpa_results_placement_spread}(c) shows that the left sensor determines the 128-bit key using less than $3k$ traces, while the right sensor retrieves less than three key bytes using $11k$ traces. 
Similarly, as shown in Fig.~\ref{fig:cpa_results_placement_spread}(d), when the number of empty slices between the LUTs and flip-flops of the AES is 2 and 3, respectively, the left sensor provides more accurate power traces compared to the right sensor. This difference is because the left sensor requires less than $2k$ traces to retrieve the 128-bit key, whereas the right sensor requires more traces to determine the 128-bit key.
When the number of empty slices between the LUTs and flip-flops of the AES is increased to 4 and 6, respectively, the right sensor is more sensitive than the left sensor, as shown in Fig.~\ref{fig:cpa_results_placement_spread}(e). The above results show the impact of inhomogeneity in the PDN structure on the sensitivity of the TDC sensors.

\textbf{Grouping flip-flops and LUTs in separate partitions.} As discussed in Section~\ref{subsec:impact_of_primitive_level_placement_of_aes_on_cpa_attack}, the flipflops and the LUTs of the AES design are grouped into separate partitions. In the AES design, 128 flip-flops and around $1.5k$ LUTs corresponding to the ciphertext. The partition block holding the flip-flops has a width of 30 slices and depth of four slices. Likewise, the block holding the LUTs has a width of 30 slices and a depth of 13 slices. There is a total of 75 empty slices between these partition blocks. As illustrated in Fig.~\ref{fig:cpa_results_placement_spread}(h), the left sensor's sensitivity is very high, as it can determine the 128-bit key with $1k$ traces. However, the right sensor could not retrieve more than two key bytes. Increasing the block width to 34 slices and reducing the block depth to three slices increases the sensitivity of both the sensors enabling the sensors to determine all the 16 key bytes, as shown in Fig.~\ref{fig:cpa_results_placement_spread}(l).

In the following section, we explain the impact of neighboring logic on the CPA attack. 

%% file: section_VI_B_defense_with_other_logic.tex
\subsection{Understanding the Impact of Neighboring Logic on PSC Attack}
\label{subsec:defense_with_other_logic}

In this section, we evaluate the resilience against PSA in a practical scenario. Generally, the crypto algorithms such as AES reside along with other designs on the FPGA rather than standalone. 
\begin{figure*}[!t]
\centering
\includegraphics[trim=0.2cm 0.2cm 0.2cm 0.2cm, clip, width=0.97\textwidth, right]{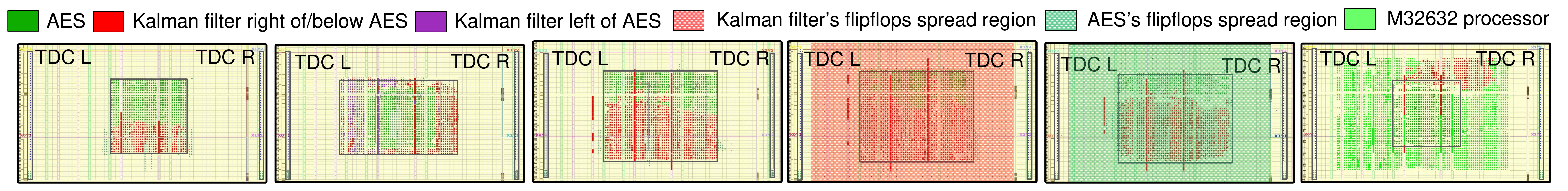}\hfill
\includegraphics[trim=0 0.24in 0 0, clip, width=\textwidth]{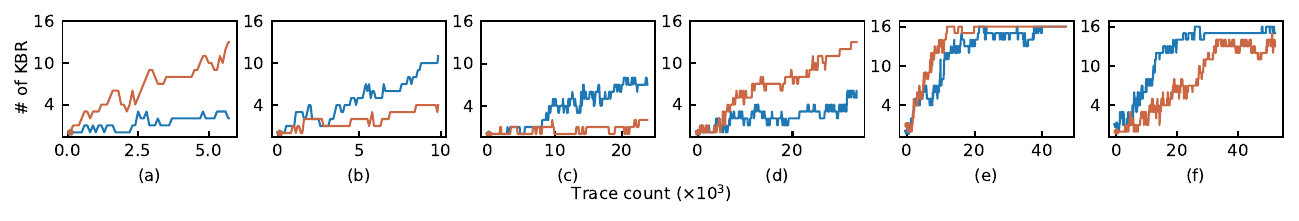}\hfill
\includegraphics[trim=0.16in 0.16in 0.16in 0.16in, clip, width=\textwidth]{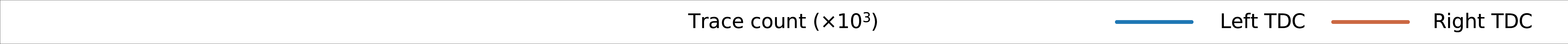}
\caption{The FPGA device view and the correlation power analysis (CPA) attack results on the different experiments with a Kalman filter and AES. (a) One Kalman filter with 16-bit input placed below the AES. (b) One Kalman filter with 16-bit input is placed on each side of AES. (c) One Kalman filter with 48-bit input placed below the AES. (d) This setup is similar to Fig.~\ref{fig:cpa_results_kalman}(c). Here, the Kalman filter's flipflops are manually spread over multiple clock regions. (e) This setup is similar to Fig.~\ref{fig:cpa_results_kalman}(c). Here, the AES's flipflops are manually spread over multiple clock regions. (f) This setup has an M32632 processor and Kalman filter along with AES. }
\label{fig:cpa_results_kalman}
\end{figure*}

\begin{figure*}[!t]
\centering
\includegraphics[trim=0.2cm 0.2cm 0.2cm 0.2cm, clip, width=0.97\textwidth, right]{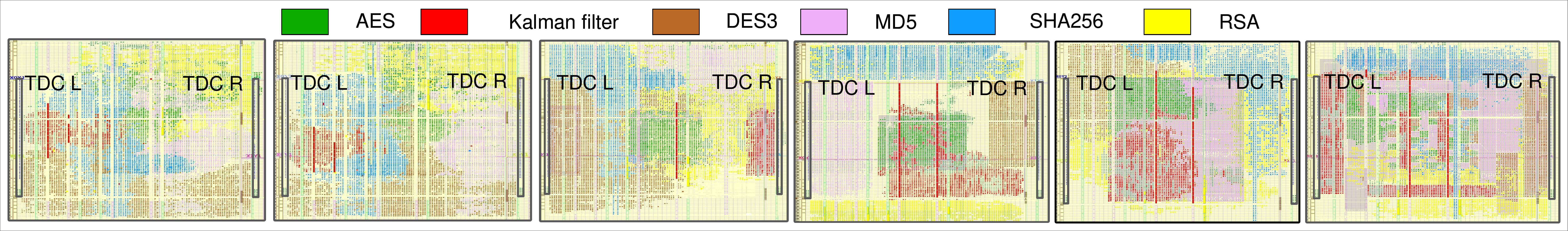}\hfill
\includegraphics[trim=0 0.24in 0 0, clip, width=\textwidth]{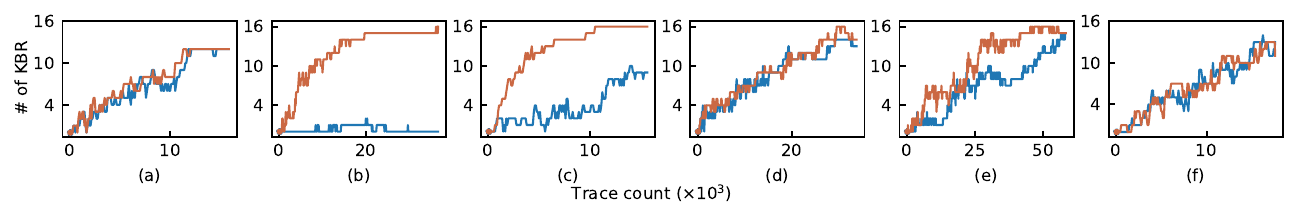}\hfill
\includegraphics[trim=0.16in 0.16in 0.16in 0.16in, clip, width=\textwidth]{legend_trace_count.pdf}
\caption{ The FPGA device view and the correlation power analysis (CPA) results on the different experiments with MIT-II CEP ~\cite{mitcep} residing along with AES. (a) The Vivado does the automatic placement of the MIT-II CEP cores. (b) The flipflops of the MIT-II CEP cores are placed in the same partition block in which AES resides. (c) The Kalman filter is placed between the right of AES and the right sensor. Similarly, DES3 is placed between the left of AES and the left sensor. (d) The activity factor of all the subblocks inside the MIT-II CEP is aligned with the AES activity factor. (e) As the Kalman filter has high activity factor compared to the other logic design in the MIT-II CEP cores, a 48-bit input Kalman filter is included in the design. (f) The defender manually does the placement, where the AES and MIT-II CEP cores are merged.}
\label{fig:cpa_results_mit_cep}
\end{figure*}

\textbf{Choice of Logic Residing Along with AES.} This crypto algorithm is used in different applications such as wireless security, processor security, file encryption, and SSL/TLS~\cite{aes_app1}. Therefore, in this work, we choose the following designs to reside along with AES: (i) the Kalman filter, (ii) the M32632 processor, and (iii) MIT-II CEP crypto cores. As filters are used predominately in wireless communications, they are chosen as one of the designs to study the resilience against PSA. An open-source processor design, M32632, is another logic chosen for this study. MIT-II CEP~\cite{mitcep} is developed to provide an open-source evaluation platform to the users to evaluate their custom tools and techniques. As this platform has a processor integrated with crypto cores, we implement that section of the CEP with the crypto cores that include MD5 (Merkle–Damg\r{a}rd), SHA256 (Secure Hash Algorithm 256), DES3 (Data Encryption Standard), and  RSA, along with the AES under protection.

\textbf{Kalman Filters.} Our first set of experiments were with Kalman filters. As discussed later in this section, the Kalman filter is either placed below or next to the AES. The FPGA device view of these placements are shown in Fig.~\ref{fig:cpa_results_kalman}(a), Fig.~\ref{fig:cpa_results_kalman}(b), and Fig.~\ref{fig:cpa_results_kalman}(c). The impact of spreading the flip-flops is studied in these experiments as well. In one build, the flip-flops in the Kalman filter are spread. While in the other, the flip-flops in the AES design are spread over more than one clock region.

\textbf{Evaluation Platform.}  As the crypto algorithms such as AES are used predominantly in processor security, we will be evaluating the resilience of the build with an open-source processor M32632 and the MIT-II CEP~\cite{mitcep}. In the case of MIT-II CEP, in this work, we implement only a part of this platform that includes all the crypto cores. We analyze the activity factor of each crypto core in the platform.  This work assumes the best-case scenario for the attacker --- he/she can place the sensors next to the boundary of the defender's logic.  Also, the impact of the activity factor over the sensors' output is maximum when they are closest to the sensors~\cite{provelengios_PDA}. Hence, in this work, the cores having high activity factors are placed close to the boundary of the allocated region, as shown in the device views in Fig.~\ref{fig:cpa_results_mit_cep}.

We shall now discuss the impact of placing these extra logic design along with AES on the MTD. Fig.~\ref{fig:cpa_results_kalman} shows the device view of different experiments with Kalman filter along with the CPA attack results for each of these experiments.

\begin{enumerate}
  \item Fig.~\ref{fig:cpa_results_kalman}(a) illustrates the device view with one Kalman filter with an input size of 16 bits. However, the CPA attack can retrieve 13 key bytes in less than $6k$ traces. A single Kalman filter is not sufficient to lower the SNR of the AES encryption. Thus, leaving it still vulnerable to PSA.
  \item Fig.~\ref{fig:cpa_results_kalman}(b) 
One Kalman filter is placed between the AES and each of the sensors. Each Kalman filter has a 16-bit input. This setup requires $10k$ power traces to determine ten key bytes, approximately twice the number of traces compared to the previous setup with one Kalman filter.
  \item Fig.~\ref{fig:cpa_results_kalman}(c) shows considerable increase in MTD. This design has only one Kalman filter; however, the input size is 48 bits.
  \item Fig.~\ref{fig:cpa_results_kalman}(d) illustrates the CPA attack results when the M32632 processor and a 16-bit Kalman filter reside along with the AES.
  \item From Section~\ref{subsubsec:primitive_placement_results} it is evident that spreading the flip-flops of the AES reduces the MTD considerably up to $1k$ traces. As the wirelength between the flip-flops and LUTs increases, the power consumed also increases. Thus, in this setup, the flip-flops of the Kalman filter are spread out. This setup increases the power consumed by the filter. Thus, we intend to increase the power consumed by the Kalman filter in the total power consumed by the defender. However, the attack results are similar to those corresponding to the setup that does not have the Kalman filter flip-flops spread out.
  \item As the wirelength between the flip-flops and LUTs increases, the power consumed also increases. \tr{In some cases, as the AES flip-flops are spread out, the sensor cannot sense the power consumption of a few of these flip-flops.} The phenomenon will increase the MTD. Thus, the AES flip-flops are spread in the setup that has the Kalman filter with 48-bit input. In this setup, the SNR of the AES operation increases due to the increased power consumption from the additional wire length. Hence, as illustrated in Fig.~\ref{fig:cpa_results_kalman}(e), the CPA attack requires less than $11k$ traces to determine all 16 key bytes.
\end{enumerate}

%% file: section_VI_C_active_fences.tex
\subsection{Active Fence Implementation~\cite{active_fences}}
\label{subsec:active_fences}

\begin{figure}[!t]
\centering
\includegraphics[trim=0.1in 0.25in 0.1in .25in, clip, width=0.48\textwidth]{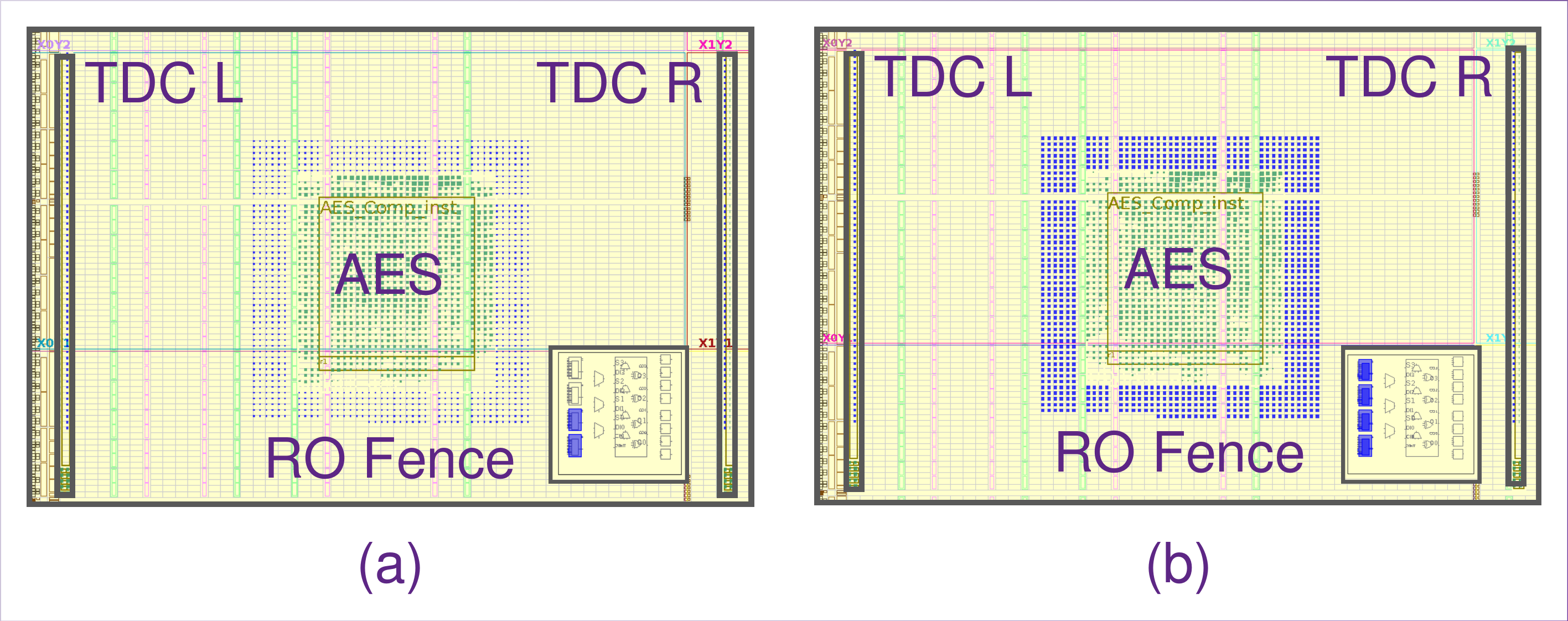}
\caption{\tp{FPGA Device view of the active fence implementation~\cite{active_fences}. Each slice in the RO fence consists of (a) two and (b) eight ROs. Each device view has an inset figure showing the LUT utilization of the slice.} }
\label{fig:active_fence}
\end{figure}

To compare our technique with the existing works, in this section, we implement the active fences defense proposed in~\cite{active_fences}. We then compare the CPA attack results of our work and this work.  The AES implemented on Zynq 706 evaluation board requires 896 slices. Hence, as per the build details in~\cite{active_fences}, 896 ROs are implemented around the AES. This implementation results in 12.5\% LUT
utilization per slice, as each slice accommodates eight LUTs. Apart from the design shared in the paper, we have tried other implementations of the ROs to check their impact on the CPA results. The other combinatorial RO design tried are as follows:

\begin{itemize}
\item 
As each slice consists of eight LUTs, two single-LUT ROs are implemented per slice, i.e.,  1792 ROs implemented in the RO fence surrounding the crypto design.
\item 
All eight LUTs implement the single-LUT RO. This build will degrade the FPGA as the fence has around $7k$ ROs, each oscillating at a frequency of around $1.4GHz$.
\item 
One RO is inferred using eight LUTs. Each of the seven LUTs infers a buffer, and the remaining one LUT infers an invertor. Finally, the eight LUTs are connected in a daisy chain fashion to implement a RO.
\item Similar to the previous design, this design has each of the seven LUTs to infer an invertor, and the remaining one LUT infers a buffer.
\end{itemize}

\begin{figure*}[!t]
\centering
\includegraphics[trim=0 0.24in 0 0, clip, width=\textwidth]{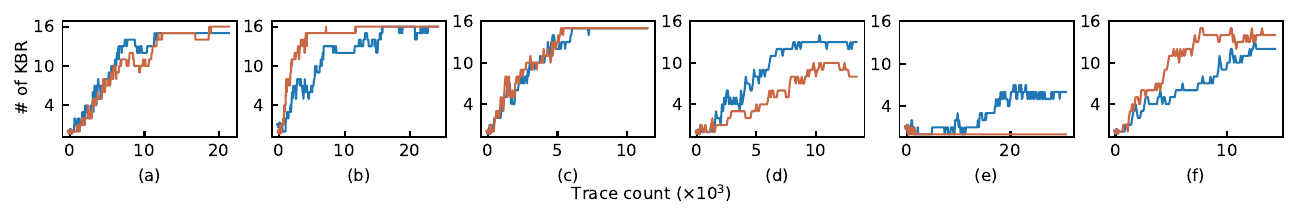}\hfill
\includegraphics[trim=0.16in 0.16in 0.16in 0.16in, clip, width=\textwidth]{legend_trace_count.pdf}
\caption{ Correlation power analysis (CPA) attack results based on the different flavors of active fences work~\cite{active_fences}. A total of 896 slices implements the ring oscillators (RO). (a) A fence separates the AES and the TDC sensor made up of 896 ROs. Each slice has one ring oscillator (RO) implemented in a single look-up table (LUT). (b)  The RO fence made up of 896 ROs surrounds the AES. The remaining builds have similar placement~\tr{as in Fig.~\ref{fig:active_fence}}. The difference is in the number of ROs in each slice and RO design. (c) The RO is implemented using seven LUTs inferred as invertors, and one LUT inferred as a buffer. (d) Each slice has two ROs, and each of these ROs are implemented using one LUT. (e) Each slice has eight ROs, and each of these ROs are implemented using one LUT.  (f) This setup has the RO design shared in Fig.~\ref{fig:old_sensor_designs}(c).}
\label{fig:cpa_results_active_fence}
\end{figure*}

\subsubsection{Impact of Active Fence on the MTD}
\label{subsubsec:active_fences_results}

The CPA attack results on the different implementations of the active fence~\cite{active_fences} shared in Section~\ref{subsec:active_fences} are explored in this section. Fig.~\ref{fig:cpa_results_active_fence}(a) plots the CPA results for the setup with RO fence only on the right of AES. Fig.~\ref{fig:active_fence}(a) illustrates the device view of this setup. The attack on the power traces from both the sensors shares similar results though the RO fence is only to the right of the AES. Placing the same number of ROs around the AES as in Fig.\ref{fig:active_fence}(b) reduces the MTD (requires less than $12k$ traces) to determine the 128-bit key. This low MTD is due to the reduced noise on each side of the AES induced by the RO. There is only one RO inferred in each slice in the builds so far. Increasing the number of ROs to two and eight per slice increase the MTD required to determine the 128-bit key, as illustrated in Fig.\ref{fig:cpa_results_active_fence}(d) and  Fig.\ref{fig:cpa_results_active_fence}(e). This increase in MTD is due to the increased activity factor as the number of ROs per slice increases.

\subsubsection{Need for RO Design Using Flip-flops}
\label{subsubsec:need_for_ff_ros}

Cloud servers such as AWS block this design from getting deployed on the F1 instance, as this technique implements combinatorial loops~\cite{aws_combo}. As ROs are used in PSC attacks~\cite{Suh18_power}, timing fault attacks~\cite{Mahmoud_timing19}, and also degrades the life of the FPGA~\cite{tahoori_voltage}, these servers thwart the loading of bitstream if they detect ROs in these bitstreams. Hence, we have also implemented the flip-flop-based ROs in the fences.  The oscillating frequency of this type of RO is $284.01MHz$, which is less than the oscillating frequency of ROs inferred using a single LUT, equal to $1.4GHz$. Thus, the activity factor of the flipflop-based ROs is less compared to the single LUT ROs. This reduction, in turn, means less dynamic power and hence, less noise due to the ROs.  Additionally, the ROs with very high oscillating frequency can have a destructive impact on the FPGA device. This impact is because the maximum clock frequency supported by the FPGA device is 800 to 900 MHz.  Hence, the routing paths in the CLBs connecting the data ports of LUTs and flip-flops can support a frequency of up to only 400 to 450MHz. Therefore, a build is generated with ROs based on flip-flops, as shown in Fig.~\ref{fig:old_sensor_designs}(c). The CPA results on the power traces collected from this build retrieve 15 out of 16 bytes using $10k$ traces, as illustrated in Fig.\ref{fig:cpa_results_active_fence}(f).

\tp{Table~\ref{tab:comparison_defense} compares the defenses to thwart PSC attack on AES on cloud FPGAs with the defense proposed in this work. As shown in this table, compared to the existing defense works, this case study based work has the following advantages: (i) study the impact of the practical environment on the CPA results, (ii) study the impact of FPGA primitive-level placements on the CPA results, (iii) does not include ROs in their defense, (iv) this defense secures against TDC-based sensors, and (v) has a higher MTD and retrieves less key bytes compared to the existing defenses. However, unlike~\cite{cpamap20}, we do not test the effectiveness of the defense at multiple locations. The onboard experimental results show that the CPA results are susceptible to the PDN design and routing apart from the placements. Hence, as part of future work, we intend to understand the PDN design and study the impact of routing on the CPA attack. Following these experiments, we will test the defense at multiple locations on the FPGA.}

\begin{table}[!b]
    \centering
    \caption{Comparison of our contributions with the existing defenses: (i) active fences~\cite{active_fences}, (ii) CPAmap~\cite{cpamap20} and (iii) votlage attack mitigation~\cite{mitigate_vol_holcomb}. CPAmap~\cite{cpamap20} requires as high as 10M traces to determine one key byte. However, this is using the active fences~\cite{active_fences} that implements combinatorial loop-based ROs. As these ROs cannot be implemented on cloud servers, we have mentioned ``not applicable'' (NA) for minimum traces for disclosure (MTD) for the CPAmap~\cite{cpamap20}.}
    \resizebox{\columnwidth}{!}{
    \label{tab:comparison_defense}
       {\tabulinesep=0.8mm
    \begin{tabu}{|p{3cm}||c|c|c|c|}
    \hline
{\bf Defense }                          & \multirow{2}{*}{\bf \cite{active_fences}}   & \multirow{2}{*}{\textbf{\cite{cpamap20}}} & \multirow{2}{*}{\textbf{\cite{mitigate_vol_holcomb}}} & {\bf Our }\\
{\bf  property}    &  &  &  & {\bf analysis}\\\hline\hline
\textbf{Practical environment}                  & \KORed                      & \KORed                   & \KORed                                 & \OKBlue \\\hline
\textbf{FPGA primitive-level placement of AES}  & \KORed                      & \KORed                   & \KORed                                 & \OKBlue  \\\hline
\textbf{Absence of ROs}                         & \KORed                      & \KORed                   & \OKBlue                                & \OKBlue  \\\hline
\textbf{Protection against TDC-based sensors}         & \OKBlue                     & \OKBlue                  & \KORed                                 & \OKBlue \\\hline
\textbf{Tested at multiple locations}           & \OKBlue                     & \OKBlue                  & \KORed                                 & \KORed \\\hline
\textbf{MTD}                                    & $10k$ & NA       & NA & $24k$ \\\hline
{\textbf{\# of key bytes retrieved}}                     & 14                &       NA        & NA &          8 \\\hline
    \end{tabu}} }  
\end{table}

%% file: section_VII_discussion_inferences.tex
\section{Inferences and Conclusion}
\label{sec:discussion_and_inferences}

\textbf{Impact of Asymmetry in PDN Structure.} Fig.~\ref{fig:cpa_results_placement_spread} and Fig.~\ref{fig:cpa_results_mit_cep} shows the change in MTD with change in the placement of the FPGA primitives corresponding to the 128-bit AES design. Though the functionality of the implemented design is same, even a small change in the placement locations causes a change in MTD. 

\begin{itemize}
\item \textbf{Spreading the Flip-flops.} As shown in Fig.~\ref{fig:cpa_results_placement_spread}(a),~\ref{fig:cpa_results_placement_spread}(b), and~\ref{fig:cpa_results_placement_spread}(c) increasing the number of empty slices between the flip-flops increases the difficulty of retrieving the AES key by the right sensor. However, the left sensor can retrive the key using 6K traces when the number of empty slices are two (minimum spread) and six (maximum spread). 
\item \textbf{Spreading the Flip-flops and LUTs.} Increasing the number of empty slices between the flip-flops and LUTs decreases and increases the difficulty of retrieving the AES key by the right and left sensors, respectively, as illustrated in Fig.~\ref{fig:cpa_results_placement_spread}(d) and ~\ref{fig:cpa_results_placement_spread}(e). 
\item \textbf{Spreading the Flip-flops and LUT Blocks.} The Fig.~\ref{fig:cpa_results_placement_spread}(h) corresponds to the CPA attack results for the build that has the partition block holding the flip-flops has a width of  30 slices and a depth of four slices.     Likewise, the block holding the LUTs has a width of 30 slices and a depth of 13 slices. There is a total of 75 empty slices between these partition blocks. Here, we could not retrieve more than two key bytes from the traces collected from the right sensor. However, increasing the width of the flip-flop and LUT block by four slices in the build makes design vulnerable CPA attack, as shown in Fig.~\ref{fig:cpa_results_placement_spread}(l). 
\item \textbf{Placing the Flip-flops of the Surrounding Logic with AES.} The Fig.~\ref{fig:cpa_results_mit_cep}(b) corresponds to the CPA attack results for the build that has the flip-flops of MIT-II CEP crypto cores reside along with the FPGA. Here, the traces collected from the left sensor does not retreive even one key byte. However, using the traces collected by the right sensor we can retrieve all 16 key bytes using 40K traces.  
\end{itemize}

\textbf{Different Clock Regions.} Fig.~\ref{fig:original_setup} shows that the left TDC and the AES share the same clock regions, X0Y0 and X0Y1; the right TDC is located in clock region X1Y0. Ideally, the left TDC must be more sensitive than the right one, determining the keys with lesser MTD. However, as shown in Fig.~\ref{fig:cpa_results_mit_cep}(b) and  Fig.~\ref{fig:cpa_results_mit_cep}(c) the right TDC is more sensitive compared to the left one. Therefore, we cannot rely on the clock regions to understand the TDC sensitivities.

%% file: section_VIII_conclusion.tex
In this work, we address the different challenges in the PSC attack and its defenses. Firstly, we enhanced the attack by manually placing the FPGA primitives corresponding to the TDC-based sensor design. This manual placement helped achieve an MTD as low as 3.8K traces to retrieve the 128-bit AES key. The impact of the junction temperature on the MTD was studied, which helped set up a repeatable attack. As a result, our attack can determine the 128-bit key every time it is launched on the FPGA. We then studied the impact of spreading the FPGA primitives corresponding to the AES on the CPA attack results. 
Finally, we also evaluated builds with additional logic residing along with AES; compared their CPA results with builds supported with active fences. Our experimental results show that the AES with additional logic having sufficient activity factor can provide the same or increased MTD compared to the build with AES surrounded by the active fences. Additionally, our defense keeps the reliability of the FPGA intact as it does not include~high-speed~combinatorial~loops.

%% file: bios.tex
\vskip -2\baselineskip plus -1fil
\begin{IEEEbiography} [{\includegraphics[width=1in,height=1.25in,clip,keepaspectratio]{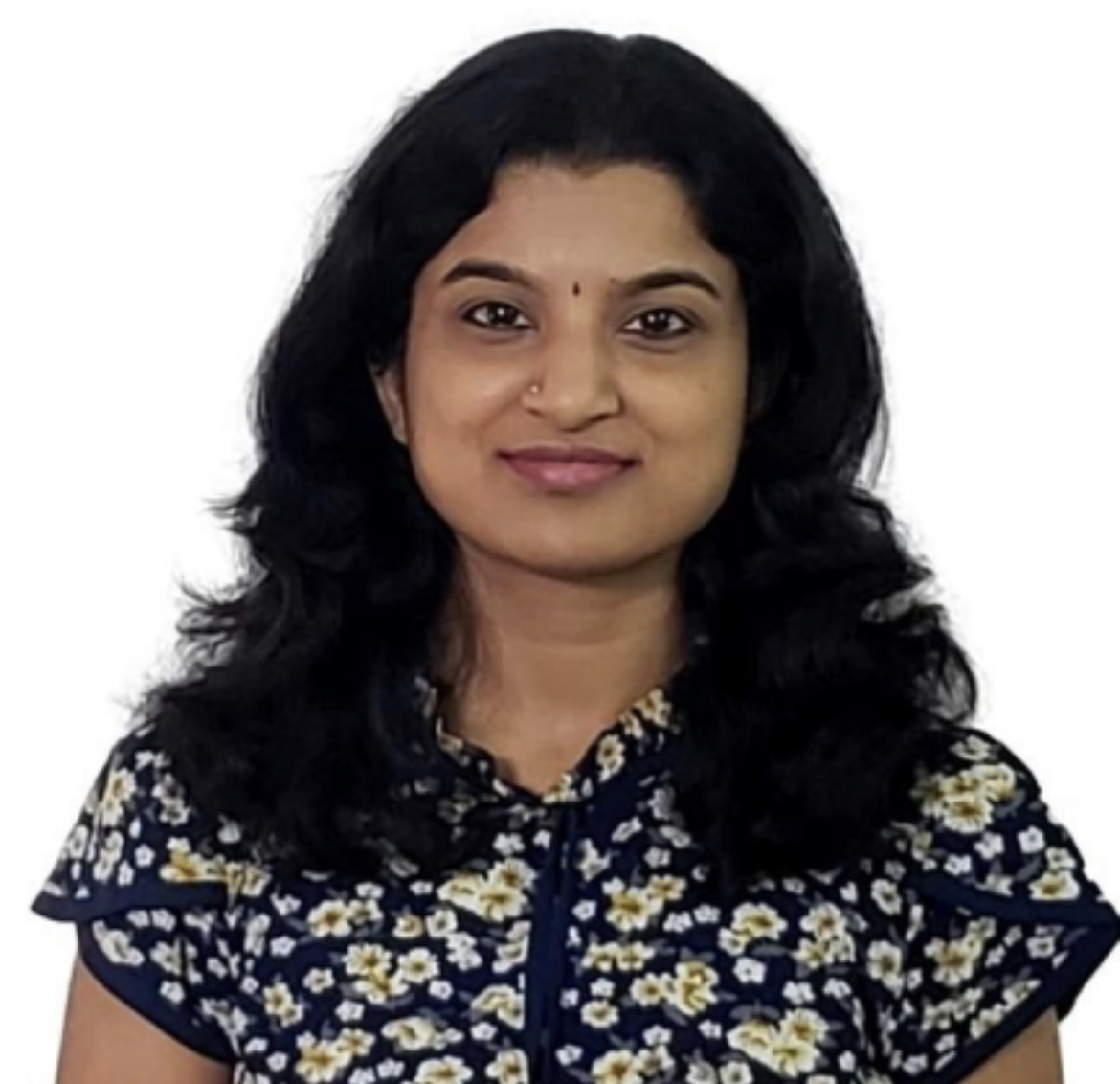}}] {Nithyashankari G. Jayasankaran} (S'18) received her B.E. degree in Electrical and Electronics Engineering from Anna University, Chennai, India, the M.S. degree in Microelectronics from Birla Institute of Science and Technology, Pilani, India, and her Ph.D. in Computer Engineering at Texas A\&M University under the guidance of Prof. Jiang Hu and Prof. JV Rajendran. Her research interest is in the hardware security domain, emphasizing analog and mixed-signal circuits security and power side-channel attacks and defenses on cloud FPGA servers. Before joining Texas A\&M, she worked on ASIC emulation, FPGA design, and board testing for base transceiver station based applications. Currently, she is working as an SoC security architect at Qualcomm, India. 
\end{IEEEbiography}

\vskip -2\baselineskip plus -1fil
\begin{IEEEbiography} [{\includegraphics[width=1in,height=1.25in,clip,keepaspectratio]{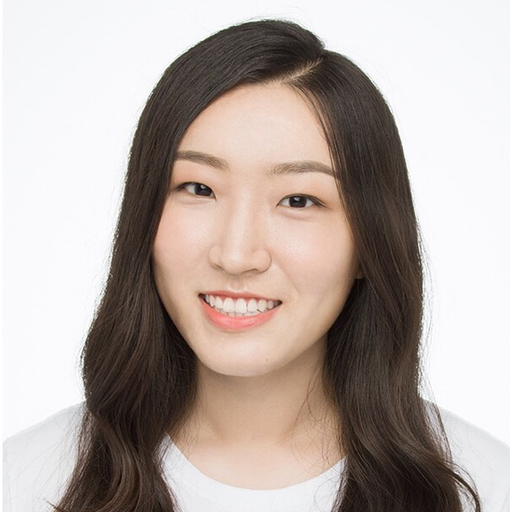}}] {Hao Guo} (S'21) received her B. Eng in Electrical Engineering and Automation from Southwest Jiaotong University, Chengdu, China, in 2017 and the M.Sc. in Electrical Engineering from University of Southern California, Los Angeles, USA, in 2019. She joined the Texas A\&M Secure and Trustworthy Hardware (SETH) Lab as a Ph.D. student in spring 2020. Her research lies in the hardware security domain focusing on FPGA security, logic locking, and applied machine learning for hardware security. She performed her summer internship at Cadence, Nanjing, China, in 2019 and worked at Qualcomm, Shanghai, China, in the R\&D department as a DFT engineer before joining SETH Lab.
\end{IEEEbiography}

\vskip -2\baselineskip plus -1fil
\begin{IEEEbiography}[{\includegraphics[width=1in,height=1.25in,clip,keepaspectratio]{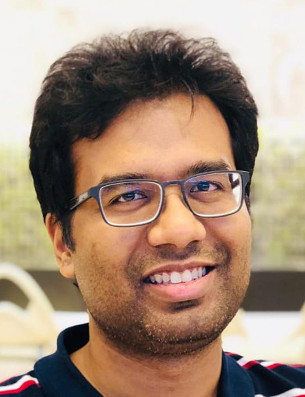}}]{Satwik Patnaik} received the B.E.\ degree in electronics and telecommunications from the University of Pune, India, the M.Tech.\ degree in computer science and engineering with a specialization in VLSI design from the Indian Institute of Information Technology and Management, Gwalior, India, and the Ph.D.\ degree in electrical engineering from Tandon School of Engineering, New York University, Brooklyn, NY, USA in September 2020.

He is currently a Postdoctoral researcher with the Department of Electrical and Computer Engineering, Texas A\&M University, College Station, TX, USA. 
His research delves into IP protection techniques, CAD frameworks for security, leveraging 3D paradigm for enhancing security, exploiting security properties of
emerging devices, applied machine learning for hardware security, and side-channel evaluation.
\end{IEEEbiography}

\vskip -2.8\baselineskip plus -1fil
\begin{IEEEbiography}
[{\includegraphics[width=1in,height=1.25in,clip,trim= 0 0 0 0.1in, keepaspectratio ]{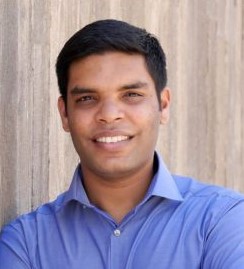}}]{Jeyavijayan (JV) Rajendran} (S'09, M'15, SM'20)  is an Assistant Professor in the Department of Electrical and Computer Engineering at the Texas A\&M University. Previously, he was an Assistant Professor at UT Dallas between 2015 and 2017. He obtained his Ph.D. degree from New York University in August 2015. His research interests include hardware security and computer security. 
His research has won the NSF CAREER Award in 2017, the ACM SIGDA Outstanding Young Faculty Award in 2019, the Intel Academic Leadership Award, and the ACM SIGDA Outstanding Ph.D. Dissertation Award in 2017.
He organizes the annual Embedded Security Challenge, a red-team/blue-team hardware security competition and has co‐founded Hack@DAC, a student security competition co-located with DAC, and FOSTER. He is a member of IEEE and ACM.
\end{IEEEbiography}

\vskip -2.4\baselineskip plus -1fil
\begin{IEEEbiography} [{\includegraphics[width=1in,height=1.25in,clip,keepaspectratio]{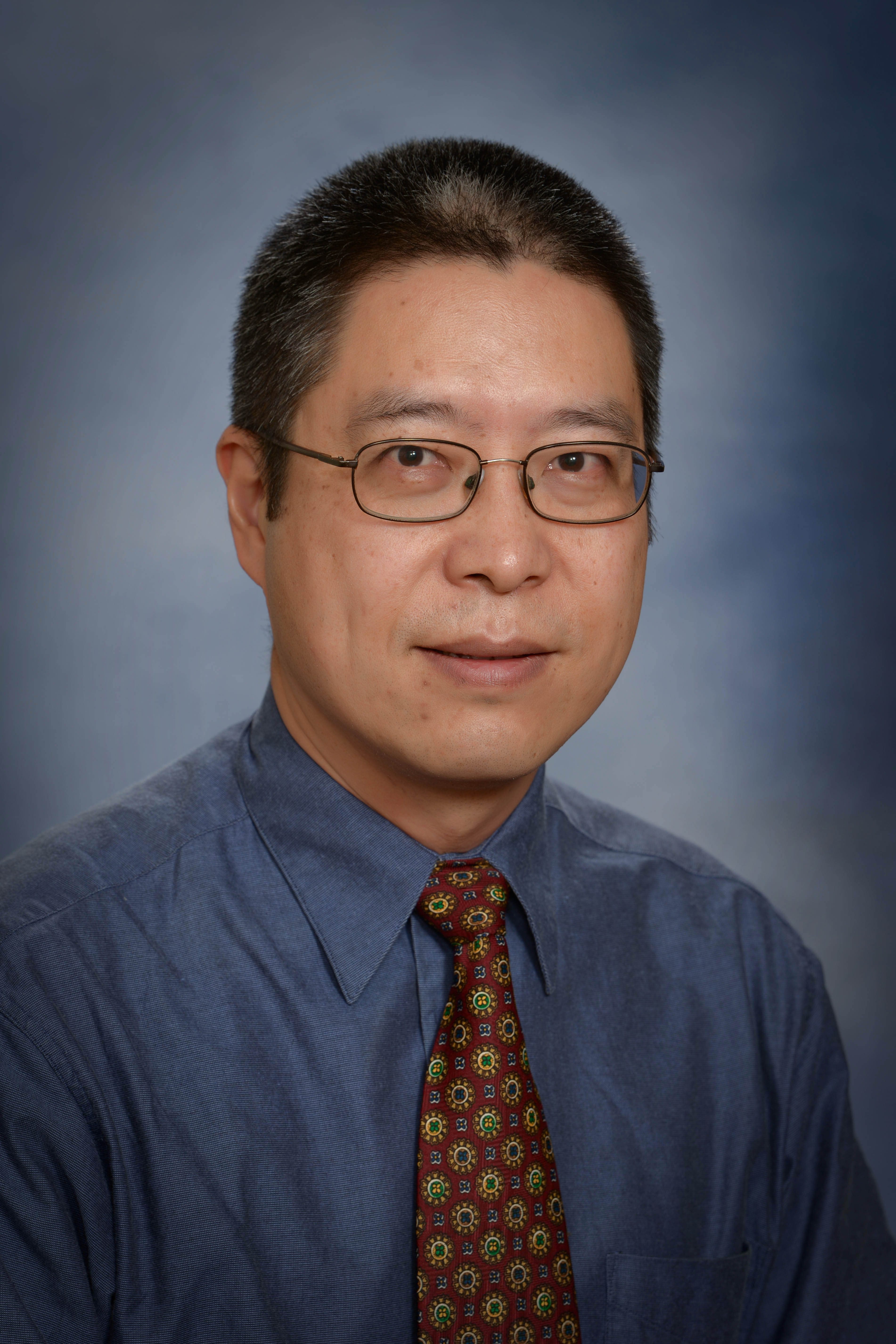}}]
{Jiang Hu} (F'16) received the Ph.D. degree from the University of Minnesota, Minneapolis, MN, USA, in 2001. He has worked with IBM Microelectronics, Armonk, NY, USA, from 2001 to 2002, and has been a Faculty Member with Texas A\&M University, College Station, TX, USA. Dr. Hu received Best Paper Awards at ACM/IEEE Design Automation Conference 2001, IEEE/ACM International Conference on Computer-Aided Design 2011, IEEE International Conference on Vehicular Electronics and Safety 2018, and IEEE/ACM International Symposium on Microarchitecture 2021. He also received the IBM Invention Achievement Award and the Humboldt Research Fellowship. He has served on the editorial boards of the IEEE Transactions on Computer-Aided Design of Integrated Circuits and Systems and the ACM Transactions on Design Automation of Electronic Systems. He was the Technical Program Chair of the ACM International Symposium on Physical Design 2011. 
\end{IEEEbiography}

%% file: main.bbl
\begin{thebibliography}{10}
\providecommand{\url}[1]{#1}
\csname url@samestyle\endcsname
\providecommand{\newblock}{\relax}
\providecommand{\bibinfo}[2]{#2}
\providecommand{\BIBentrySTDinterwordspacing}{\spaceskip=0pt\relax}
\providecommand{\BIBentryALTinterwordstretchfactor}{4}
\providecommand{\BIBentryALTinterwordspacing}{\spaceskip=\fontdimen2\font plus
\BIBentryALTinterwordstretchfactor\fontdimen3\font minus
  \fontdimen4\font\relax}
\providecommand{\BIBforeignlanguage}[2]{{%
\expandafter\ifx\csname l@#1\endcsname\relax
\typeout{** WARNING: IEEEtran.bst: No hyphenation pattern has been}%
\typeout{** loaded for the language `#1'. Using the pattern for}%
\typeout{** the default language instead.}%
\else
\language=\csname l@#1\endcsname
\fi
#2}}
\providecommand{\BIBdecl}{\relax}
\BIBdecl

\bibitem{amazon_f1}
Amazon, ``{Amazon EC2 F1 Instance},''
  \url{https://aws.amazon.com/ec2/instance-types/f1/}, 2016, {Last accessed on
  09/05/2021}.

\bibitem{micro_azure}
Microsoft, ``{Microsoft Azure},'' \url{https://azure.microsoft.com/en-gb/},
  2021, {Last accessed on 09/05/2021}.

\bibitem{ibm_cloud_fpga}
IBM, ``{CloudFPGA},'' \url{https://www.zurich.ibm.com/cci/cloudFPGA/}, 2020,
  {Last accessed on 09/05/2021}.

\bibitem{tacc}
U.~of~Texas~at Austin, ``{Texas Advanced Computing Center},''
  \url{https://www.tacc.utexas.edu/}, 2020, {Last accessed on 09/05/2021}.

\bibitem{cloudfpga_server}
C.~Jin, V.~Gohil, R.~Karri, and J.~Rajendran, ``{Security of Cloud FPGAs: A
  Survey},'' \emph{arXiv}, 2020.

\bibitem{avnet_fpga}
Avnet, ``{Where is FPGA in Cloud Computing Today?}''
  \url{https://www.avnet.com/wps/portal/apac/resources/article/where-is-fpga-in-cloud-computing-today/},
  2019, {Last accessed on 09/05/2021}.

\bibitem{mt_uses}
Z.~Istv\'{a}n, G.~Alonso, and A.~Singla, ``{Providing Multi-tenant Services
  with FPGAs: Case Study on a Key-Value Store},'' \emph{IEEE International
  Conference on Field Programmable Logic and Applications}, pp. 119--1195,
  2018.

\bibitem{Suh18_power}
M.~{Zhao} and G.~E. {Suh}, ``{FPGA-Based Remote Power Side-Channel Attacks},''
  \emph{IEEE Symposium on Security and Privacy}, pp. 229--244, 2018.

\bibitem{cld_vulnerable}
O.~{Glamo\v{c}anin}, L.~{Coulon}, F.~{Regazzoni}, and M.~{Stojilovi\'c}, ``{Are
  Cloud FPGAs Really Vulnerable to Power Analysis Attacks?}'' \emph{IEEE/ACM
  Design, Automation \& Test in Europe}, pp. 1007--1010, 2020.

\bibitem{pda20}
G.~Provelengios, D.~Holcomb, and R.~Tessier, ``{Power Distribution Attacks in
  Multitenant FPGAs},'' \emph{IEEE Transactions on Very Large Scale Integration
  Systems}, vol.~28, no.~12, pp. 2685--2698, 2020.

\bibitem{inside_job}
F.~Schellenberg, D.~R. Gnad, A.~Moradi, and M.~B. Tahoori, ``{An Inside Job:
  Remote Power Analysis Attacks on FPGAs},'' \emph{Design, Automation \& Test
  in Europe}, 2021.

\bibitem{Ramesh18}
C.~{Ramesh}, S.~B. {Patil}, S.~N. {Dhanuskodi}, G.~{Provelengios},
  S.~{Pillement}, D.~{Holcomb}, and R.~{Tessier}, ``{FPGA Side Channel Attacks
  without Physical Access},'' \emph{Annual International Symposium on
  Field-Programmable Custom Computing Machines}, pp. 45--52, 2018.

\bibitem{remote_inter_tahoori}
F.~Schellenberg, D.~R. Gnad, A.~Moradi, and M.~B. Tahoori, ``{Remote Inter-Chip
  Power Analysis Side-Channel Attacks at Board-Level},'' \emph{{IEEE/ACM
  International Conference on Computer-Aided Design}}, pp. 1--7, 2018.

\bibitem{tahoori_voltage}
D.~R.~E. {Gnad}, F.~{Oboril}, and M.~B. {Tahoori}, ``{Voltage Drop-based Fault
  Attacks on FPGAs Using Valid Bitstreams},'' \emph{IEEE International
  Conference on Field Programmable Logic and Applications}, pp. 1--7, 2017.

\bibitem{Krautter_Gnad_Tahoori_2018}
J.~Krautter, D.~R.~E. Gnad, and M.~B. Tahoori, ``{FPGAhammer: Remote Voltage
  Fault Attacks on Shared FPGAs, suitable for DFA on AES},'' \emph{IACR
  Transactions on Cryptographic Hardware and Embedded Systems}, vol. 2018,
  no.~3, pp. 44--68, 2018.

\bibitem{provelengios_PDA}
G.~{Provelengios}, D.~{Holcomb}, and R.~{Tessier}, ``{Characterizing Power
  Distribution Attacks in Multi-User FPGA Environments},'' \emph{IEEE
  International Conference on Field Programmable Logic and Applications}, pp.
  194--201, 2019.

\bibitem{Mahmoud_timing19}
D.~{Mahmoud} and M.~{Stojilovi\'c}, ``{Timing Violation Induced Faults in
  Multi-Tenant FPGAs},'' \emph{IEEE/ACM Design, Automation \& Test in Europe},
  pp. 1745--1750, 2019.

\bibitem{osc19}
T.~Sugawara, K.~Sakiyama, S.~Nashimoto, D.~Suzuki, and T.~Nagatsuka,
  ``{Oscillator without a Combinatorial Loop and Its Threat to FPGA in Data
  Centre},'' \emph{Electronics Letter}, vol.~55, no.~11, pp. 640--642, 2020.

\bibitem{giechaskiel19}
I.~Giechaskiel, K.~Eguro, and K.~B. Rasmussen, ``{Leakier Wires: Exploiting
  FPGA Long Wires for Covert- and Side-Channel Attacks},'' \emph{ACM
  Transactions on Reconfigurable Technology and Systems}, vol.~12, no.~3, 2019.

\bibitem{gie_iccd19}
I.~Giechaskiel, K.~Rasmussen, and J.~Szefer, ``{Reading Between the Dies:
  Cross-SLR Covert Channels on Multi-Tenant Cloud FPGAs},'' \emph{IEEE
  International Conference on Computer Design}, pp. 1--10, 2019.

\bibitem{giechaskiel-a}
------, ``{CAPSULe: Cross-FPGA Covert-Channel Attacks through Power Supply Unit
  Leakage},'' \emph{IEEE Symposium on Security and Privacy}, pp. 909--922,
  2020.

\bibitem{gie_asiaccs18}
I.~Giechaskiel, K.~Rasmussen, and K.~Eguro, ``{Leaky Wires: Information Leakage
  and Covert Communication Between FPGA Long Wires},'' \emph{ACM Asia
  Conference on Computer and Communications Security}, pp. 18--27, 2018.

\bibitem{active_fences}
J.~Krautter, D.~R. Gnad, F.~Schellenberg, A.~Moradi, and M.~B. Tahoori,
  ``{Active Fences against Voltage-based Side Channels in Multi-Tenant
  FPGAs},'' \emph{IEEE/ACM International Conference on Computer-Aided Design},
  pp. 1--8, 2019.

\bibitem{cpamap20}
J.~Krautter, D.~R.~E. Gnad, and M.~B. Tahoori, ``{CPAmap: On the Complexity of
  Secure FPGA Virtualization, Multi-Tenancy, and Physical Design},'' \emph{IACR
  Transactions on Cryptographic Hardware and Embedded Systems}, vol. 2020,
  no.~3, pp. 121--146, 2020.

\bibitem{tahoori_fpga_voltage}
D.~R.~E. Gnad, C.~D.~K. Nguyen, S.~H. Gillani, and M.~B. Tahoori,
  ``{Voltage-based Covert Channels in Multi-Tenant FPGAs},'' Cryptology ePrint
  Archive, Report 2019/1394, 2019, \url{https://eprint.iacr.org/2019/1394}.

\bibitem{temporal_sca19}
S.~Tian and J.~Szefer, ``{Temporal Thermal Covert Channels in Cloud FPGAs},''
  \emph{ACM/SIGDA International Symposium on Field-Programmable Gate Arrays},
  p. 298–303, 2019.

\bibitem{mitigate_vol_attacks21}
J.~Krautter, D.~R.~E. Gnad, and M.~B. Tahoori, ``{Mitigating Electrical-Level
  Attacks towards Secure Multi-Tenant FPGAs in the Cloud},'' \emph{ACM
  Transactions on Reconfigurable Technology and System}, vol.~12, no.~3, 2019.

\bibitem{mitcep}
{Massachusetts Institute of Technology}, ``{Common Evaluation Platform},''
  \url{https://github.com/mit-ll/CEP}, 2021, {Last accessed on 09/05/2021}.

\bibitem{psa_dpa_attack}
P.~Kocher, J.~Jaffe, and B.~Jun, ``{Differential Power Analysis},''
  \emph{Advances in Cryptology}, pp. 388--397, 1999.

\bibitem{pant_des}
S.~Pant, ``{Design and Analysis of Power Distribution Networks in VLSI
  Circuits},'' \emph{Ph.D. dissertation, The University of Michigan}, 2008.

\bibitem{giechaskiel19_fpl}
I.~{Giechaskiel}, K.~B. {Rasmussen}, and J.~{Szefer}, ``{Measuring Long Wire
  Leakage with Ring Oscillators in Cloud FPGAs},'' \emph{IEEE International
  Conference on Field Programmable Logic and Applications}, pp. 45--50, 2019.

\bibitem{tahoori18_electrical}
D.~R.~E. {Gnad}, S.~{Rapp}, J.~{Krautter}, and M.~B. {Tahoori}, ``{Checking for
  Electrical Level Security Threats in Bitstreams for Multi-tenant FPGAs},''
  \emph{IEEE International Conference on Field-Programmable Technology}, pp.
  286--289, 2018.

\bibitem{aws_combo}
Amazon, ``{Combinatorial Loops in F1 FPGA},''
  \url{https://forums.aws.amazon.com/thread.jspa?threadID=264137}, 2017, {Last
  accessed on 09/05/2021}.

\bibitem{Zick2013SensingNV}
K.~M. Zick, M.~Srivastav, W.~Zhang, and M.~C. French, ``{Sensing
  Nanosecond-Scale Voltage Attacks and Natural Transients in FPGAs},''
  \emph{{ACM/SIGDA International Symposium on Field-Programmable Gate Arrays}},
  2013.

\bibitem{tdc_vernier_ro}
K.~{Cui} and X.~{Li}, ``{A High-Linearity Vernier Time-to-Digital Converter on
  FPGAs with Improved Resolution Using Bidirectional-Operating Vernier Delay
  Lines},'' \emph{IEEE Transactions on Instrumentation and Measurement}, pp.
  1--1, 2019.

\bibitem{Xia_tdc_6_6}
H.~Xia, G.~Cao, and N.~Dong, ``{A 6.6 ps RMS Resolution Time-to-Digital
  Converter Using Interleaved Sampling Method in a 29 nm FPGA},'' \emph{Review
  of Scientific Instruments}, vol.~90, no.~4, p. 044706, 2019.

\bibitem{wang_tdc_11}
J.~{Wang}, S.~{Liu}, L.~{Zhao}, X.~{Hu}, and Q.~{An}, ``{The 10-ps Multitime
  Measurements Averaging TDC Implemented in an FPGA},'' \emph{IEEE Transactions
  on Nuclear Science}, vol.~58, no.~4, pp. 2011--2018, 2011.

\bibitem{sensors_study}
S.~Moini, X.~Li, P.~Stanwicks, G.~Provelengios, W.~Burleson, R.~Tessier, and
  D.~Holcomb, ``{Understanding and Comparing the Capabilities of On-Chip
  Voltage Sensors against Remote Power Attacks on FPGAs},'' \emph{IEEE
  International Midwest Symposium on Circuits and Systems}, pp. 941--944, 2020.

\bibitem{cpa_attack}
E.~Brier, C.~Clavier, and F.~Olivier, ``{Correlation Power Analysis with a
  Leakage Model},'' \emph{{Cryptographic Hardware and Embedded Systems}}, pp.
  16--29, 2004.

\bibitem{mitigate_vol_holcomb}
G.~Provelengios, D.~Holcomb, and R.~Tessier, ``{Mitigating Voltage Attacks in
  Multi-Tenant FPGAs},'' \emph{ACM Transactions on Reconfigurable Technology
  and Systems}, 2021.

\bibitem{aes_code}
{Tohoku University}, ``{Cryptographic Hardware Prject},''
  \url{http://www.aoki.ecei.tohoku.ac.jp/crypto/web/cores.html}, 2007, {Last
  accessed on 09/05/2021}.

\bibitem{vio_xilinx}
Xilinx, ``{Virtual Input/Output v3.0},''
  \url{https://www.xilinx.com/support/documentation/ip_documentation/vio/v3_0/pg159-vio.pdf},
  2018, {Last accessed on 09/05/2021}.

\bibitem{cpa_fei}
{NUEESS Lab Northeastern University}, ``{Side Channel Analysis Library},''
  \url{https://bit.ly/2Y6k0ZD}, 2013, {Last accessed on 09/05/2021}.

\bibitem{xdc_xilinx}
Xilinx, ``{Vivado Design Suite Properties Reference Guide},''
  \url{https://bit.ly/2XeABu1}, 2020, {Last accessed on 09/09/2021}.

\bibitem{xilinx_power_methodology}
------, ``{Power Methodology Guide},''
  \url{https://www.xilinx.com/support/documentation/sw_manuals/xilinx14_7/ug786_PowerMethodology.pdf},
  2013, {Last accessed on 09/05/2021}.

\bibitem{aes_app1}
R.~Thomas, ``{Advanced Encryption Standard (AES): What It Is and How It
  Works},'' \url{https://bit.ly/3zOeQzd}, 2020, {Last accessed on 09/05/2021}.

\end{thebibliography}
